\documentclass[pre,twocolumn,showpacs,superscriptaddress,preprintnumbers,floatfix]{revtex4-1}

\usepackage{etex}
\usepackage{ifpdf}
\usepackage{hyperref}
\usepackage{dcolumn}
\usepackage{url}
\usepackage{amsmath}
\usepackage{amscd}
\usepackage{amsfonts}
\usepackage{amssymb}
\usepackage{amsthm}
\usepackage{xfrac}
\usepackage{braket}
\usepackage{bm}   
\usepackage{bbm}
\usepackage{verbatim}
\usepackage{stmaryrd}
\usepackage{xcolor}
\usepackage{setspace}
\usepackage{tikz}
\usetikzlibrary{arrows}
\usetikzlibrary{automata}
\usetikzlibrary{positioning}
\usetikzlibrary{plotmarks}
\usepackage{pgfplots}
\pgfplotsset{compat=newest}
\usepackage{floatrow}

\tikzstyle{vaucanson}=[
  node distance=2.5cm,
  bend angle=15,
  auto,
  every loop/.style={},
  every edge/.style={->,draw=black,line width=1.2,>=latex,shorten <=1pt,shorten >=1pt},
  every state/.style={draw=black,line width=2,font=\large},
  loop right/.style={right,out=22,in=-22,loop},
  loop above/.style={above,out=112,in=68,loop},
  loop left/.style={left,out=202,in=158,loop},
  loop below/.style={below,out=292,in=248,loop},
  loop above right/.style={right,out=67,in=23,loop},
  loop above left/.style={left,out=157,in=113,loop},
  loop below left/.style={left,out=247,in=203,loop},
  loop below right/.style={right,out=337,in=293,loop},
]

\theoremstyle{plain}    \newtheorem{Lem}{Lemma}
\theoremstyle{plain}    \newtheorem*{ProLem}{Proof}
\theoremstyle{plain}    
\theoremstyle{plain}    
\theoremstyle{plain}    \newtheorem{The}{Theorem}
\theoremstyle{plain}    \newtheorem*{ProThe}{Proof}
\theoremstyle{plain}    \newtheorem{Prop}{Proposition}
\theoremstyle{plain}    \newtheorem*{ProProp}{Proof}
\theoremstyle{plain}    
\theoremstyle{plain}    
\theoremstyle{plain}    
\theoremstyle{plain}    
\theoremstyle{plain}

\DeclareMathOperator*{\argmin}{argmin}

\newcommand{\eM}     {\mbox{$\epsilon$-machine}}
\newcommand{\eMs}    {\mbox{$\epsilon$-machines}}

\newcommand{\Process}{\mathcal{P}}

\newcommand{\MeasAlphabet}  {\mathcal{A}}
\newcommand{\SampleSpace} { \AllPasts }
\newcommand{\MeasSymbol}   { {X} }
\newcommand{\meassymbol}   { {x} }

\newcommand{\Past} { \smash{\overleftarrow {\MeasSymbol}} }
\newcommand{\past} { {\smash{\overleftarrow {\meassymbol}}} }

\newcommand{\Future}   { \smash{\overrightarrow{\MeasSymbol}} }
\newcommand{\future}   { \smash{\overrightarrow{\meassymbol}} }

\newcommand{\AllPasts}      { \smash{\overleftarrow{ {\rm {\bf \MeasSymbol}} } } }

\newcommand{\rep}    { r }
\newcommand{\Rep}    { \mathcal{R} }
\newcommand{\repa}    { \widehat{r} }
\newcommand{\Repa}    { \widehat{\mathcal{R}} }
\newcommand{\repb}    { \widetilde{r} }
\newcommand{\Repb}    { \widetilde{\mathcal{R}} }

\newcommand{\CausalState}   { \mathcal{S} }

\newcommand{\causalstate}   { \sigma }
\newcommand{\CausalStateSet}    { \bm{\CausalState} }
\newcommand{\AlternateState}    { \mathcal{R} }

\newcommand{\AlternateStateSet} { \bm{\AlternateState} }

\newcommand{\Prob}      {\Pr} 

\newcommand{\Cmu}       {C_\mu}
\newcommand{\hmu}       {h_\mu}
\newcommand{\EE}        {{\bf E}}

\newcommand{\ProcessAlphabet}   {\MeasAlphabet}

\newcommand{\forward}{+}
\newcommand{\reverse}{-}
\newcommand{\forwardreverse}{\pm} 

\newcommand{\FutureCausalState} { {\CausalState}^{\forward} }
\newcommand{\futurecausalstate} { \sigma^{\forward} }

\newcommand{\PastCausalState}   { {\CausalState}^{\reverse} }
\newcommand{\pastcausalstate}   { \sigma^{\reverse} }

\newcommand{\FutureCmu} { C_\mu^{\forward} }
\newcommand{\PastCmu}   { C_\mu^{\reverse} }

\newcommand{\lastindex}[2]{
  \edef\tempa{0}
  \edef\tempb{#2}
  \ifx\tempa\tempb
    \edef\tempc{#1}
  \else
    \edef\tempa{0}
    \edef\tempb{#1}
    \ifx\tempa\tempb
      \edef\tempc{#2}
    \else
      \edef\tempc{#1+#2}
    \fi
  \fi
  \tempc
}

\newcommand{\sigmamu}{\sigma_\mu}

\newcommand{\I}{\mathbf{I}}

\newcommand{\CSjoint}[1][,]{
   \edef\tempa{:}
   \edef\tempb{#1}
   \ifx\tempa\tempb
      \ensuremath{\FutureCausalState\!#1\PastCausalState}
   \else
      \ensuremath{\FutureCausalState#1\PastCausalState}
   \fi
}

\newif\ifpm
\edef\tempa{\forwardreverse}
\edef\tempb{\pm}
\ifx\tempa\tempb
   \pmfalse
\else
   \pmtrue
\fi

\renewcommand{\H}{\operatorname{H}}
\renewcommand{\I}{\operatorname{I}}

\colorlet {R_color}    {blue}
\colorlet {k_color}    {black!30!green}

\newcommand{\DKL} { {D_{\textrm{KL}} } }

\def\clap#1{\hbox to 0pt{\hss#1\hss}}

\begin{document}

\title{Circumventing the Curse of Dimensionality in Prediction:\\
\vspace{0.05in}
Causal Rate-Distortion for\\
Infinite-Order Markov Processes}

\author{Sarah Marzen}
\email{smarzen@berkeley.edu}
\affiliation{Department of Physics,\\
Redwood Center for Theoretical Neuroscience\\
University of California at Berkeley, Berkeley, CA 94720-5800}

\author{James P. Crutchfield}
\email{chaos@ucdavis.edu}
\affiliation{Complexity Sciences Center and Department of Physics\\
University of California at Davis, One Shields Avenue, Davis, CA 95616}

\date{\today}
\bibliographystyle{unsrt}

\begin{abstract}
{
Predictive rate-distortion analysis suffers from the curse of dimensionality:
clustering arbitrarily long pasts to retain information about arbitrarily long
futures requires resources that typically grow exponentially with length. The
challenge is compounded for infinite-order Markov processes, since conditioning
on finite sequences cannot capture all of their past dependencies. Spectral
arguments show that algorithms which cluster finite-length sequences fail
dramatically when the underlying process has long-range temporal correlations
and can fail even for processes generated by finite-memory hidden Markov
models. We circumvent the curse of dimensionality in rate-distortion analysis
of infinite-order processes by casting predictive rate-distortion objective
functions in terms of the forward- and reverse-time causal states of
computational mechanics. Examples demonstrate that the resulting causal
rate-distortion theory substantially improves current predictive
rate-distortion analyses.
}

\vspace{0.2in}

\end{abstract}

\preprint{Santa Fe Institute Working Paper 14-12-XXX}
\preprint{arxiv.org:1412.XXXX [cond-mat.stat-mech]}

\maketitle


\setstretch{1.1}

\newcommand{\Abet}{\ProcessAlphabet}
\newcommand{\MS}{\MeasSymbol}
\newcommand{\ms}{\meassymbol}
\newcommand{\SSet}{\CausalStateSet}
\newcommand{\St}{\CausalState}
\newcommand{\st}{\causalstate}
\newcommand{\FSt}{\FutureCausalState}
\newcommand{\fst}{\futurecausalstate}
\newcommand{\FCmu}{\FutureCmu}
\newcommand{\PCmu}{\PastCmu}
\newcommand{\PSt}{\PastCausalState}
\newcommand{\pst}{\pastcausalstate}
\newcommand{\MxSt}{\AlternateState}
\newcommand{\MxSSet}{\AlternateStateSet}
\newcommand{\mxst}{\mu}
\newcommand{\mxstt}[1]{\mu_{#1}}
\newcommand{\StartMS}{\bra{\delta_\pi}}

\newcommand{\CodeRate}  { \I [\Past;\AlternateState] }
\newcommand{\Shielding} { \I [\Past;\Future | \AlternateState] }
\newcommand{\StateFutI} { \I [\AlternateState | \Future ] }

\section{Introduction}

Biological organisms and engineered devices are often required to predict the
future of their environment either for survival or performance. Absent side
information about the environment that is inherited or hardwired, their only
guide to the future is the past. One strategy for adapting to environmental
challenges, then, is to memorize as much of the past as possible---a strategy
that ultimately fails, even for simple stochastic environments, due to the
exponential growth in required resources.


One way to circumvent resource limitations is to identify maximally predictive
features in a time series, coarse-graining pasts into partitions with
equivalent conditional probability distributions over futures. These partitions
are minimal sufficient statistics for prediction, known as the
\emph{forward-time causal states} $\St^+$ of \emph{computational mechanics},
and storing them costs on average $\FCmu = \H[\FSt]$ bits \cite{Crut88a,
Shal98a}. For processes with finite $\FCmu$
\cite{Varn14a,Crut12a} identifying the causal states obviates storing an
exponentially growing number of sequences. However, for many processes, perhaps
most \cite{Crut92c, Uppe97a}, $\FCmu$ is infinite and so storing the causal
states themselves exceeds the capacity of any learning strategy.

As such, one asks for approximate, lossy features that predict the future as
well as possible given resource constraints. Shannon introduced
\emph{rate-distortion theory} to analyze such trade-offs
\cite{Shan48a,Shan59a}, encoding an information source so that ``the maximum
possible signaling rate is obtained without exceeding the tolerable distortion
level''.  When applied to prediction, rate-distortion theory provides a
principled framework for calculating the function delineating achievable from
unachievable predictive distortion for a given amount of memory. Such
rate-distortion functions are used to test, for instance, whether or not a
biological sensory system extracts lossy predictive features \cite{Palm13a}. In
other applications, optimal codebooks identify useful features for
understanding and building approximate predictive models of complex datasets
\cite{Creutzig08,Creutzig09,singh2004predictive,singh2003learning,dutech2013partially}
and define natural predictive macrostates of a stochastic process
\cite{Stil07a,Stil07b}.

Unfortunately, current methods for calculating rate-distortion
functions for prediction require clustering arbitrarily long
pasts to obtain information about arbitrarily long futures, incurring the
very resource limitations one hoped to avoid. They can be avoided
for some classes of simple process, when there is an analytic expression for
the joint past-future probability distribution \cite{chechik2005information,
Creutzig09, Rey2012}. In practice, though, one compresses finite-length pasts
to retain information about finite-length futures \cite{Stil07a, Stil07b}. This
will yield reasonable estimates of predictive rate-distortion functions at
sufficient lengths, but how long is long enough?

To address this practical problem and, more generally, to circumvent resource
limitations, we identify a new relationship between predictive rate-distortion
theory and causal states. This gives an alternative theory and class of algorithms for
calculating predictive rate-distortion functions, when a maximally predictive
model is identified first. Revisiting previous results \cite{Stil07a,Stil07b},
the alternative demonstrates that identifying a maximally predictive
model first dramatically improves rate-distortion analysis for even the
simplest stochastic processes. A natural hierarchy of phase transitions
associated with discovering new predictive features emerges as a function of
approximation error, paralleling those described in Ref. \cite{Rose94a}. As an
illustration, we calculate a predictive ``hierarchy'' for a process generated
by a complex chaotic dynamical system.

Section \ref{sec:Background} reviews computational mechanics and predictive
rate-distortion theory. Section \ref{sec:CurseofDimensionality} describes how
current predictive rate-distortion algorithms encounter the curse of
dimensionality. Section \ref{sec:CRD} introduces a theorem that reformulates
predictive rate-distortion analysis in terms of forward- and reverse-time
causal states. Section \ref{sec:Examples} then describes a new algorithm for
computing lossy causal states and illustrates its performance on
infinite-order Markov processes. Section \ref{sec:Conclusion} summarizes
outstanding issues, desirable extensions, and future applications.

\section{Background}
\label{sec:Background}

When an information source's entropy rate falls below a channel's capacity,
Shannon's Second Coding Theorem says that there exists an encoding of the
source messages such that the information can be transmitted error-free, even
over a noisy channel. What happens, though, when the source rate is above this
error-free regime? This is what Shannon solved by introducing
rate-distortion theory \cite{Shan48a,Shan59a}.

\subsection{Problem Statement}
\label{sec:Problem}

Our view is that, for natural systems, the above-capacity regime is
disproportionately more common and important than the original error-free
coding with which Shannon and followers started. This is certainly the
circumstance in which most of measurement science finds itself. Instrumentation
almost never exactly captures an experimental system's states exactly. One can
argue that this is even guaranteed by quantum uncertainty relations. Similarly
for biological life and engineered adaptive systems, their sensory apparatus
does not capture all of an environment's information and organization. Indeed,
in many cases they cannot due to sensory limitations. Even without such
limitations, moreover, they may not have adequate representational capacity.

And, perhaps more profoundly, one does not want to do perfectly, as the
``stammering grandeur" of Irenio Funes reminds us \cite[Funes The
Memorious]{Borg62a}. Said positively, summarizing sensory information not only
helps reduce demands on memory, but also the computational complexity of
downstream perceptual processing, cognition, and acting. For instance, much
effort has been focused on determining memory and the ability to reproduce a
given time series \cite{Vita93a}, but that memory may only be important to the
extent that it affects the ability to predict the future; e.g., see Ref.
\cite{Crut92c, Bial01a}. A decision that is adaptive for an organism can be quite
simple, requiring only a coarse sketch of the environmental state: Individual
\textit{Dictyostelium discoideum} slime mold cells track only the concentration gradient of cyclic-AMP when moving to organize into a collective fruiting bud \cite{Devr89a}. Moreover, many human perceptual models are rooted in
identifying informative environmental features \cite{Ande91a}. Experiments
suggest there are constraints on the total number of features that humans can
identify \cite{Gers13a}---a psychological variant of the coding problem
tackled by rate-distortion theory, but in a different asymptotic limit. We are interested, therefore, as others have been, in identifying lossy causal states.

First, we review causal states. Second, we review several information measures
of stochastic processes. These, finally, lead us to describe what we mean by
lossy causal states. The following assumes familiarity with information theory
at the level of Ref. \cite{Cove06a,Yeun08a}, information theory for complex
processes at the level of Refs. \cite{Crut01a,Jame11a}, and computational
mechanics at the level of Ref. \cite{Crut12a}.

\subsection{Processes and Their Causal States}
\label{sec:Background_CM}

When predicting a system the main object is the \emph{process} $\Process$ it
generates: the list of all of a system's behaviors or realizations $\{ \ldots
\ms_{-2}, \ms_{-1}, \ms_{0}, \ms_{1}, \ldots \}$ as specified by their joint
probabilities $\Prob(\ldots \MS_{-2}, \MS_{-1}, \MS_{0}, \MS_{1}, \ldots)$. We
denote a contiguous chain of random variables as $\MS_{0:\ell} = \MS_0 \MS_1
\cdots \MS_{\ell-1}$. Left indices are inclusive; right, exclusive. We suppress
indices that are infinite.  In this setting, the \emph{present}
$\MS_{t:t+\ell}$ is the length-$\ell$ chain beginning at $t$, the \emph{past}
is the chain $\MS_{:t} = \ldots \MS_{t-2} \MS_{t-1}$ leading up the present,
and the \emph{future} is the chain following the present $\MS_{t+\ell:} =
\MS_{t+\ell+1} \MS_{t+\ell+2} \cdots$. When being more expository, we use arrow
notation; for example, for the past $\Past = \MS_{:0}$ and future $\Future =
\MS_{0:}$. We will refer on occasion to the space $\AllPasts$ of all pasts.
Finally, we assume a process is ergodic and stationary---$\Prob(\MS_{0:\ell}) =
\Prob(\MS_{t:\ell+t})$ for all $t \in \mathbb{Z}$---and the measurement symbols
$\ms_t$ range over a finite alphabet: $\ms \in \Abet$. We make no assumption
that the symbols represent the system's states---they are at best an indirect
reflection of an internal Markov mechanism. That is, the process a system
generates is a \emph{hidden Markov process} \cite{Ephr02a}.

Forward-time causal states $\St^+$ are minimal sufficient statistics for
predicting a process's future \cite{Crut88a,Shal98a}. This follows
from their definition as sets of pasts grouped by the equivalence
relation $\sim^+$:
\begin{align}
\ms_{:0} \sim^+ & \ms_{:0}' \nonumber \\
  & \Leftrightarrow
  \Prob (\MS_{0:}|\MS_{:0}=\ms_{:0}) = \Prob(\MS_{0:}|\MS_{:0}=\ms_{:0}')
  ~.
\label{eq:PredEquivReln}
\end{align}
As a shorthand, we denote a cluster of pasts so defined, a \emph{causal state},
as $\st^+ \in \St^+$. We implement Eq. (\ref{eq:PredEquivReln}) via the
\emph{causal state map}: $\st^+ = \epsilon^+ (\past)$. Through it, each state
$\st^+$ inherits a probability $\pi(\st^+)$ from the process's probability over
pasts $\Prob(\MS_{:0})$. The forward-time \emph{statistical complexity} is
defined as the Shannon entropy of the probability distribution over
forward-time causal states \cite{Crut88a}:
\begin{equation}
\Cmu^+ = \H[\St^+]
  ~.
\end{equation}

\newcommand{\cs} {\causalstate}

A generative model---the process's \emph{\eM}---is built out of the causal
states by endowing the state set with a transition dynamic:
\begin{align*}
T_{\cs\cs'}^{x} = \Prob(\St_{t+1}^+=\st',\MS_t = \ms|\St_t^+=\st)
  ~,
\end{align*}
matrices that give the probability of generating the next symbol $\ms_t$ and
ending in the next state $\cs_{t+1}$, if starting in state $\cs_t$. (Since
output symbols are generated during transitions there is, in effect, a half
time-step difference in index. We suppress notating this.) For a discrete-time,
discrete-alphabet process, the \eM\ is its minimal unifilar Hidden Markov Model
(HMM) \cite{Crut88a,Shal98a}. (For general background on HMMs see Refs.
\cite{Paz71a,Rabi86a,Rabi89a}. For a mathematical development of \eMs\ see
Ref.  \cite{Lohr2009models}.) Note that the causal-state set of a process
generated by even a finite HMM can be finite, countable, or uncountable.
\emph{Minimality} can be defined by either the smallest number of causal states
or the smallest statistical complexity $\Cmu$ \cite{Shal98a}.
\emph{Unifilarity} is a constraint on the transition matrices such that the
next state $\st_{t+1}$ is determined by knowing the current state $\st_t$ and
the next symbol $\ms_t$.

A similar equivalence relation can be applied to find minimal sufficient
statistics for retrodiction \cite{Crut08a}. Futures are grouped together if
they have equivalent conditional probability distributions over pasts:
\begin{align}
\ms_{0:} \sim^- & \ms_{0:}' \nonumber \\
  & \Leftrightarrow
  \Prob(\MS_{:0}|\MS_{0:}=\ms_{0:}) = \Prob(\MS_{:0}|\MS_{0:}=\ms_{0:}')
  ~.
\label{eq:CausalEquiv}
\end{align}
A cluster of futures---a \emph{reverse-time causal state}---defined by $\sim^-$
is denoted $\st^-\in\St^-$. Again, each $\st^-$ inherits a probability
$\pi(\st^-)$ from the probability over futures $\Prob(\MS_{0:})$. And, the
\emph{reverse-time statistical complexity} is the Shannon entropy of the
probability distribution over reverse-time causal states:
\begin{equation}
\Cmu^- = \H[\St^-]
  ~.
\end{equation}
In general, the forward- and reverse-time statistical complexities are not equal
\cite{Crut08a, Elli11a}. That is, different amounts of information must be
stored from the past (future) to predict (retrodict). Their difference $\Xi =
C_{\mu}^+-C_{\mu}^-$ is a process's \emph{causal irreversibility} and it
reflects this statistical asymmetry.

\begin{figure}[htp]
\centering
\includegraphics[width=0.5\columnwidth]{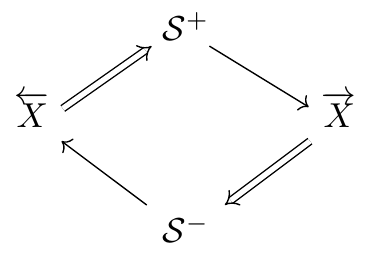}
\caption{Computational Mechanics Markov Loop \cite{Crut08a,Crut08b}:
  Markov chain relationships between the past, future, and
  causal states $\FSt$ and $\PSt$.
  Relations: $\Rightarrow$ denotes ``function of'', and $\rightarrow$
  denotes being part of Markov chain.
  }
\label{fig:Intuition1}
\end{figure}

Here, the most important aspect of forward- and reverse-time causal states are
that they ``shield'' the past and future from one another. That is:
\begin{align*}
\Prob(\Past,\Future|\St^+) & = \Prob(\Past|\St^+)\Prob(\Future|\St^+)
  \text{ and } \\
  \Prob(\Past,\Future|\St^-) & = \Prob(\Past|\St^-)\Prob(\Future|\St^-)
  ~,
\end{align*}
even though $\St^+$ and $\St^-$ are functions of $\Past$ and $\Future$,
respectively. Figure \ref{fig:Intuition1} illustrates shielding the future
$\Future$ from the past $\Past$, given by the forward causal states $\St^+$,
and the past from the future, given by the reverse causal states $\St^-$. The
result is a series of Markov chains forming a Markov loop that illustrates the
relationship between prediction and retrodiction.

\subsection{Information Measures}

To measure a process's asymptotic per-symbol uncertainty one uses the
Shannon entropy rate:
\begin{align*}
\hmu = \lim_{\ell \to \infty} \frac{\H(\ell)}{\ell}
  ~,
\end{align*}
when the limit exists and where $\H(\ell) = -\sum_{w \in \MeasAlphabet^\ell}
\Prob(w) \log_2 \Prob(w)$ is the \emph{block entropy}. $\hmu$ measures the rate
at which a stochastic process generates information. From standard
informational identities, one sees that the entropy rate is also given by the
conditional entropy:
\begin{align}
\hmu = \lim_{\ell \to \infty} \H[\MS_0|\MS_{-\ell:0}] ~.
\label{eq:EntropyRateConditional}
\end{align}
This form makes transparent its interpretation as the residual uncertainty in a
measurement given the infinite past. As such, it is often employed as a measure
of a process's degree of unpredictability. The maximum amount of information in
the future that is \emph{predictable} from the past (or vice versa) is the
\emph{excess entropy}:
\begin{align*}
\EE & = \I[\MS_{:0};\MS_{0:}] ~.
\end{align*}
It is symmetric in time and a lower bound on the stored information:
$\EE \leq \Cmu$.

More generally, Shannon's various information quantities---entropy, conditional
entropy, mutual information, and the like---when applied to processes are
functions of the joint distributions $\Prob(\MS_{0:\ell})$. Importantly, they
define an algebra of information measures for a given set of random variables
\cite{Yeun08a}. Reference \cite{Jame11a} used this to show that the past and
future partition the single-measurement entropy $\H(\MS_0)$ into several
distinct measure-theoretic atoms. These are useful in particular to answer the
question posed by the Introduction, how long is long enough to capture all of a
process's correlations? For this, we use the amount of predictable information
not captured by the length-$\ell$ present:
\begin{align}
\sigmamu(\ell) = \I[\MS_{:0};\MS_{\ell:}|\MS_{0:\ell}]
  ~.
\end{align}
This is the \emph{elusive information} \cite{Ara14a}, which measures the amount
of past-future correlation not contained in the present. It is a decreasing
function of the present's length $\ell$. It is nonzero if a process
\emph{necessarily} has hidden states and is therefore quite sensitive to how
a system's internal state space is observed or coarse grained. For example, an order-$R$ Markov process has
$\sigmamu(\ell) = 0$ for $\ell \geq R$. In this case, $R$-blocks serve as a
process's effective states, rendering the past and future
independent---shielding them from each other. Note, however, that in the
space of processes generated by finite HMMs, infinite-order Markov processes
dominate overwhelmingly \cite{Jame10a}. For these, $\sigmamu(\ell)$ vanishes
only asymptotically. The excess entropy itself is lower-bounded by the elusive
information: $\sigmamu(\ell) \leq \EE$. In fact, $\sigmamu(0) = \EE$.

Finally, the following refers to finite-length variants of causal states and
finite-length estimates of statistical complexity and $\EE$. For example, the
latter is given by:
\begin{equation}
\EE(M,N) = \\I[\MS_{-M:0};\MS_{0:N}]
  ~.
\end{equation}
If $\EE$ is finite, then $\EE = \lim_{M,N\rightarrow\infty} \EE(M,N)$.
Processes generated by finite-state HMMs have finite $\EE$---they are
\emph{finitary} processes \cite{Crut01a}.
When $\EE$ is infinite, then the way in which $\EE(M,N)$ diverges is one measure of
a process' complexity \cite{Crut89e,Bial00a,Crut01a}. An analogous, finite
past-future $(M,N)$-parametrized equivalence relation leads to finite-length
causal states---$\FSt_{M,N}$ and $\PSt_{M,N}$---and statistical
complexities---$\FutureCmu(M,N) = \H[\FSt_{M,N}]$ and $\PastCmu(M,N) =
\H[\PSt_{M,N}]$.

The quality of a process's approximation can be monitored by the convergence error $\EE - \EE(M,N)$, which is controlled by the elusive information $\sigmamu(\ell)$. To see this, we apply the mutual information chain rule repeatedly:
\begin{align*}
\EE & = \I[\MS_{:0};\MS_{0:}] \\
  & = \I[\MS_{:0};\MS_{0:N-1}] + \sigmamu(N) \\
  & = \EE(M,N) + \I[\MS_{:-M-1};\MS_{0:N-1}|\MS_{-M-1:0}] + \sigmamu(N)
  ~.
\end{align*}
The last mutual information is difficult to interpret, but easy to bound:
\begin{align*}
\I[\MS_{:-M-1};\MS_{0:N-1} & |\MS_{-M-1:0}] \\
  & \leq \I[\MS_{:-M-1};\MS_{0:}|\MS_{-M-1:0}] \\
  & = \sigmamu(M)
  ~,
\end{align*}
And so, the convergence error is upper-bounded by the elusive information:
\begin{align}
0 \leq \EE - \EE(M,N) \leq \sigmamu(N) + \sigmamu(M)
  ~.
\label{eq:OCF_Bound}
\end{align}

\subsection{Lossy Causal States}

\newcommand{\Capacity} { \mathcal{C} }
\newcommand{\Codebook} { \mathbb{C} }

Lossy causal states are naturally defined via \emph{predictive rate-distortion}
(PRD) or its information-theoretic instantiations, \emph{optimal causal
inference} (OCI) \cite{Stil07a, Stil07b} and the \emph{past-future information
bottleneck} (PFIB) \cite{Creutzig09}. This section briefly reviews
rate-distortion theory, but interested readers should refer to
Refs.~\cite{Shan48a,Shan59a,Gray90b} or Ref. \cite[Ch. 8]{Yeun08a} for detailed
expositions. The presentation here is adapted to serve our focus on prediction.

\begin{figure}[htp]
\centering
\includegraphics[width=\columnwidth]{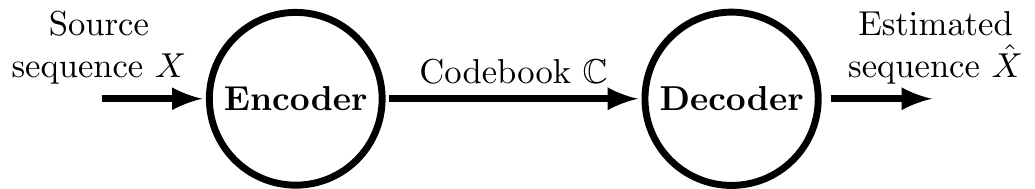}
\includegraphics[width=0.5\columnwidth]{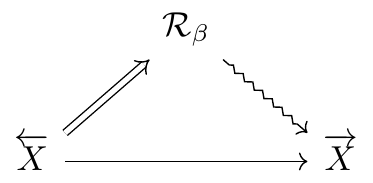}
\caption{(top) Rate-distortion setting: Using a codebook an encoder maps
  input sequences into codewords. A decoder then estimates the source
  sequence from the codewords.
  (bottom) Optimal Causal Inference \cite{Stil07a, Stil07b}:
  Markov chain relationships between the past, future, and lossy
  predictive features $\Rep_\beta$.
  Relations: $\Rightarrow$ denotes ``function of'', $\rightarrow$
  denotes a Markov chain, and $Y \rightsquigarrow Z$
  indicates that random variable $Z$ is the ``relevant'' variable for $Y$.
  The latter does not imply a Markov chain relationship.
  }
\label{fig:Intuition2}
\end{figure}

Figure~\ref{fig:Intuition2}(top) shows the rate-distortion theory setting of
Ref. \cite{Yeun08a} that combines the encoder and noisy channel used in
Shannon's communication channel model \cite{Shan48a, Shan59a} into a single
encoder. This is often a more appropriate framing for biological information
processing \cite{Palm13a} where a sensory system (e.g., retina) both distorts
the input signal (e.g., natural scenery) and transmits codewords that convey
information about the input signal to the next information processing post
(e.g., the lateral geniculate nucleus).

In this framing, an information source generates $n$ successive symbols, a
\emph{word} $\ms_{0:n} \in \MeasAlphabet^n$. The word is presented to the
encoder, which outputs one of $M$ codewords $\{ \rep_i : i = 1, \ldots M \}$.
The collection $\Codebook = \{(\ms_{0:n},\rep_i) : i = 1, \ldots, M\}$, mapping
words to codewords, is the channel \emph{codebook}. Since sending one of $M$
codewords requires $\log_2 M$ bits, the \emph{code rate} $R(\Codebook) =
n^{-1} \log M$ is the number of communicated bits per source symbol.

Given a codeword $\rep_i$, the decoder makes its best estimate
$\widehat{\ms}_{0:n} \in \MeasAlphabet^n$ as to the original source word
$\ms_{0:n}$. Each estimate is evaluated using a distortion measure
$d(\ms_{0:n},\widehat{\ms}_{0:n}) \geq 0$, that quantifies the error
between the given codeword $\widehat{\ms}_{0:n}$ and the true word $\ms_{0:n}$.
Over long periods, the estimates lead to an expected distortion:
\begin{align*}
\mathbb{E} d(\ms,\widehat{\ms})
  = \lim_{n\rightarrow\infty}
  \frac{\mathbb{E} d(\ms_{0:n},\widehat{\ms}_{0:n})}{n}
  ~,
\end{align*}
that depends on the codebook $\Codebook$ and its associated code rate
$R(\Codebook)$. We assume that the distortion measure is \emph{normal}:
$\min_{\{\widehat{\ms}_{0:n}\}} d(\ms_{0:n},\widehat{\ms}_{0:n}) = 0$.

There is a natural trade-off between the code rate and the expected distortion:
the smaller the desired expected distortion, the larger the required code rate
to achieve it. To achieve no expected distortion, $\mathbb{E}
d(\ms,\widehat{\ms}) = 0$, requires some minimal code rate $R_{max}$, which
depends on the distortion measure. For instance, when \textit{reconstructing} a
given time series as well as possible, $R_{max}$ is the process's entropy rate
$\hmu$ \cite{Stei96a}. When the information source consists of successive
semi-infinite pasts $\Past$ of a time series, then for many
\textit{prediction}-related distortion measures, $R_{max}$ is the forward-time
statistical complexity $\FCmu$ \cite{Shal98a, Stil07a, Stil07b}. More
generally, when the distortion measure is an informational distortion of a
source $X$ with respect to some ``relevant variable'' $Y$ \cite{Tish00a,
slonim2002information}, then $R_{max}$ is the entropy of the inherited
probability distribution of the minimal sufficient statistics of $X$ with
respect to $Y$ \cite{shamir2010learning}.

A channel's \emph{capacity} $\Capacity$ is the largest information transmission
rate it can sustain over all possible information sources $X$:
\begin{align*}
\Capacity = \sup_{\Prob(X)} \I[X;Y] ~,
\end{align*}
where $Y$ is the channel's output process. If the encoder's capacity is large
enough---$\Capacity \geq R_{max}$---then there exists a codebook such that the
decoder can reconstruct the input sequence with arbitrarily small probability
of error. If the encoder's capacity is not large
enough---$\Capacity < R_{max}$---however, it cannot. There is an irreducible
positive error rate. As the Introduction noted, for PRD applications one is in
this regime when $\FCmu$ is infinite \cite{Uppe97a, Crut92c}.

In this positive error-rate regime, we ask for the \emph{rate-distortion function}:
\begin{align}
R(D) = \inf_{\mathbb{E} d(\ms,\widehat{\ms}) \leq D} R(\Codebook)
  ~.
\label{eq:RD}
\end{align}
For simplicity, we typically limit ourselves to single-symbol distortion
measures, so that the distortion between decoded and input word is:
\begin{align*}
d(\ms_{0:n},\widehat{\ms}_{0:n}) = \sum_{i=0}^{n-1} d(\ms_i,\widehat{\ms}_i)
  ~.
\end{align*}
Equation (\ref{eq:RD}) is a difficult optimization, given that to determine
$R(D)$ for each $D$ requires enumerating a combinatorially large space of codebooks. Instead, one views the information source as a random variable $\MS$
with realizations $\ms$ and the (potentially stochastic) codebook's output as a
random variable $\Rep = \Codebook(X)$ with realizations $\rep \in \{1, \ldots,
M\}$.

Then, according to the Rate-Distortion Theorem, one has:
\begin{align*}
R(D) = \min_{\langle d(\ms,\rep)\rangle_{\MS,\Rep} \leq D} \I[\MS;\Rep]
  ~.
\end{align*}
One can solve numerically for $R(D)$ using annealing to find a
$\Prob(\Rep|\MS)$ minimizing the objective function:
\begin{align*}
\mathcal{L}_{\beta} = \I[\MS;\Rep]
  + \beta \langle d(\ms,\rep)\rangle_{\MS,\Rep}
  ~.
\end{align*}
As $\beta$ varies, one obtains different rates $R_{\beta} = \I[\MS;\Rep]$ and
expected distortions $D_{\beta} = \langle d(\ms,\rep)\rangle_{\MS,\Rep}$. In
this way, one traces out the rate-distortion function $R(D)$ parametrically.

Note that, in the preceding, $\MS$ was not a measurement symbol of a stochastic
process, as it was in Sec. \ref{sec:Background_CM}. In the present context,
$\MS$ is the random variable denoting the output of any information source,
which could be, but need not be, a string of symbols from a stochastic process.

A natural question arises, which distortion measure should one use?
Historically, it was chosen to be a more or less familiar statistic, such as
the mean-squared error and the like. More recently, though, information
measures have been used, mirroring the definition of the code rate in the
objective function. For example, Ref.~\cite{Tish00a} posited that one should
retain some aspect $\Rep$ of the input $\MS$ that is related to another
``relevant'' random variable $Y$. This is tantamount to assuming a Markov chain
relationship: $\Rep \to X \to Y$. A natural distortion measure then is
\cite{harremoes2007information}:
\begin{align*}
d(\ms,\rep) = \DKL[\Prob(Y|\MS=\ms)||\Prob(Y|\Rep=\rep)]
  ~,
\end{align*}
where $\DKL(P||Q)$ is the relative entropy between distributions $P$ and $Q$.
Though, one might consider others based on the distance between the conditional
probability distributions $\Prob(Y|\MS=\ms)$ and $\Prob(Y|\Rep=\rep)$. It is
straightforward to show that:
\begin{align*}
\langle \DKL[\Prob(Y|\MS=\ms)||\Prob(Y|\Rep=\rep)]\rangle = \I[\MS;Y|\Rep]
  ~,
\end{align*}
since the Markov chain gives $\Prob(Y|X,\Rep) = \Prob(Y|X)$.
And, since $\I[\MS;Y|\Rep] = \I[\MS;Y]-\I[\Rep;Y]$ due to the same Markov
chain assumption, we define the
\textit{information function}:
\begin{align*}
R(I_0) = \min_{\I[\Rep;Y]\geq I_0} \I[\MS;\Rep] ~,
\end{align*}
with objection function:
\begin{align*}
\mathcal{L}_{\beta} = \I[\Rep;Y] - \beta^{-1} \I[\MS;\Rep]
  ~.
\end{align*}
Finding the $\Prob(\Rep|\MS)$ that maximize $\mathcal{L}_{\beta}$ is the basis
for the \emph{information bottleneck} (IB) method \cite{Tish00a,
slonim2002information}.

Since our focus is prediction, we have already decided that the information
source is a process's past $\Past$ with realizations $\past$ and the relevant
variable is its future $\Future$.
We have the Markov chain $\Rep \to \Past \to \Future$. (See Fig. \ref{fig:Intuition2}(bottom).)
And, our distortion measures have the form: 
\begin{align*}
d(\past,\rep) = d(\Prob(\Future|\Past=\past),\Prob(\Future|\Rep=\rep))
  ~.
\end{align*}
The predictive rate-distortion function is then:
\begin{align*}
R(D) = \min_{\langle d(\past,\rep)\rangle_{\Past,\Rep} \leq D} \I[\Rep;\Past]
  ~.
\end{align*}
Determining the optimal $\Prob(\Rep|\Past)$ that achieve these limits is
\emph{predictive rate distortion} (PRD). If we restrict to informational
distortions, such as $\I[\Past;\Future|\Rep]$, we find the associated information function is:
\begin{align}
R(I_0) = \min_{\I[\Rep;\Future]\geq I_0} \I[\Rep;\Past]
  	~,
\label{eq:InfoCurve}
\end{align}
and its accompanying objective function is:
\begin{align}
\mathcal{L}_{\beta} = \I[\Rep;\Future] - \beta^{-1} \I[\Past;\Rep]
  ~.
\label{eq:OCIObjective}
\end{align}
Calculating the $\Prob(\Rep|\Past)$ that maximize Eq. (\ref{eq:OCIObjective})
constitutes \emph{optimal causal inference} (OCI) \cite{Stil07a,Stil07b} or,
for linear dynamical systems, the \emph{past-future information bottleneck}
\cite{Creutzig09}. We refer to these methods generally as the \emph{predictive
information bottleneck} (PIB), emphasizing that an information distortion measure
for PRD has been chosen.

This choice of method name does lead to
confusion since the \emph{recursive information bottleneck} (RIB) introduced in
Ref. \cite{Stil13a} is an information bottleneck approach to predictive
inference that does not take the form of Eq.~(\ref{eq:OCIObjective}). However,
RIB is a departure from the original IB framework since its objective function
explicitly infers lossy machines rather than lossy statistics \cite{Shal99a}.

Generally, rate-distortion analysis reveals how a process's structure
(captured in the associated codebooks) varies under coarse-graining, with the
shape of the rate-distortion functions identifying those regimes in which smaller codebooks are good approximations. To implement this analysis,
one graphs a process's \emph{information function} $R(I_0)$ and its accompanying
\emph{feature curve} $(\beta,R_{\beta})$, corresponding to optimizing
$\mathcal{L}_{\beta}$ above. Again, previous results established
that the zero-distortion predictive features are a process's causal states and
so the maximal $R(I_0) = \Cmu^+$ \cite{Stil07a, Stil07b} and this code rate occurs at an $I_0 = \EE$ \cite{Crut08a,Crut08b}.

Figure \ref{fig:Intuition2}(bottom) illustrates how the approximately
predictive states $\AlternateState_\beta$ are soft clusters of the past $\Past$
constrained by predicting the future $\Future$ at a fidelity controlled by
$\beta$. The new notation introduced there to indicate which random variable is
``relevant'' will be useful for explaining Thm. \ref{the:CRD} via
Fig.~\ref{fig:MarkovChains_CRD}(b).

\section{Curse of Dimensionality in Prediction}
\label{sec:CurseofDimensionality}

Let's consider the performance of any PRD algorithm that clusters pasts of
length $M$ to retain information about futures of length $N$. In the lossless
limit, when these algorithms work, they find features that capture
$\I[\MS_{-M:0};\MS_{0:N}] = \EE(M,N)$ of the total predictable information
$\EE$ at a coding cost of $\Cmu^{+}(M,N)$. As $M, N \to \infty$, they should
recover the forward-time causal states with predictability $\EE$ and coding
cost $\Cmu^+$. Increasing $M$ and $N$ come with an associated computational
cost, though: storing the joint probability distribution
$\Prob(\MS_{-M:0},\MS_{0:N})$ of past and future finite-length trajectories
requires storing $|\MeasAlphabet|^{M+N}$ probabilities.

More to the point, applying PRD algorithms at small distortions
requires storing and manipulating a matrix of dimension $|\MeasAlphabet|^{M} \times |\MeasAlphabet|^N$.
This leads to obvious practical limitations---the \emph{curse of dimensionality
for prediction}. For example, current computing is limited to matrices of size
$10^5 \times 10^5$ or less, thereby restricting rate-distortion analyses to
$M,N \leq \log_{|\MeasAlphabet|} 10^5$. (Recall that this is an overestimate,
since the sparseness of the
sequence distribution is determined by a process's topological entropy rate.) And
so, even for a binary process, when $|\MeasAlphabet| = 2$, one is practically
limited to $M,N \leq 16$. Notably, $M,N\leq 5$ are more often used in practice
\cite{Creutzig08,Stil07a,Stil07b,gueguen2007image, gueguen2006analysis}.
Finally, note that these estimates do not account for the computational costs
of managing numerical inaccuracies when measuring or manipulating the
vanishingly small sequence probabilities that occur at large $M$ and $N$.

These constraints compete against achieving good approximations of the
information function: we require that $\EE - \EE(M,N)$ be small. Otherwise,
approximate information functions provide a rather weak lower bound on the true
information function for larger code rates. This has been noted before in other contexts, when
approximating non-Gaussian distributions as Gaussians leads to significant
underestimates of information functions \cite{Rey2012}. This calls for an
independent calibration for convergence. We
address this by calculating $\EE - \EE(M,N)$ in terms of the transition matrix
$W$ of a process' mixed-state presentation. When $W$ is diagonalizable with
eigenvalues $\{\lambda_i\}$, Ref. \cite{Ara14b} provides the closed-form
expression:
\begin{align}
& \EE - \EE(M,N) = \nonumber \\
  & \sum_{i:\lambda_i\neq 1} \frac{\lambda_i^{M}+\lambda_i^{N+1}
  - \lambda_i^{M+N+1}}{1-\lambda_i}
  \langle \delta_{\pi}|W_{\lambda_i}|H(W^{\MeasAlphabet})\rangle
  ,
\label{eq:IpredMN}
\end{align}
where $\langle \delta_{\pi}|W_{\lambda_i}|H(W^{\MeasAlphabet})\rangle$ is a dot
product between the eigenvector $\langle \delta_{\pi}|W_{\lambda_i}$
corresponding to eigenvalue $\lambda_i$ and a vector $H(W^{\MeasAlphabet})$
of transition uncertainties out of each mixed state. Here, $\pi$ is the
stationary state distribution and $W_{\lambda_i}$ is the projection
operator associated with $\lambda_i$. When $W$'s spectral gap $\gamma =
1-\max_{i:\lambda_i\neq 1} |\lambda_i|$ is small, then $\EE(M,N)$ necessarily
asymptotes more slowly to $\EE$. When $\gamma$ is small, then (loosely
speaking) we need $M,N \sim \log_{1-\gamma} \frac{\epsilon}{\gamma}$, where
$\epsilon$ is of order $\EE - \EE(M,N)$.

\begin{figure}[h]
\includegraphics[width=0.8\columnwidth]{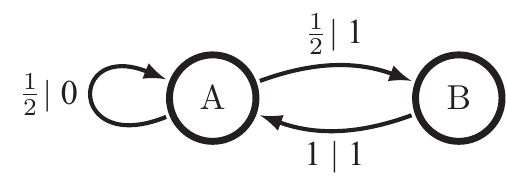}
\includegraphics[width=\textwidth]{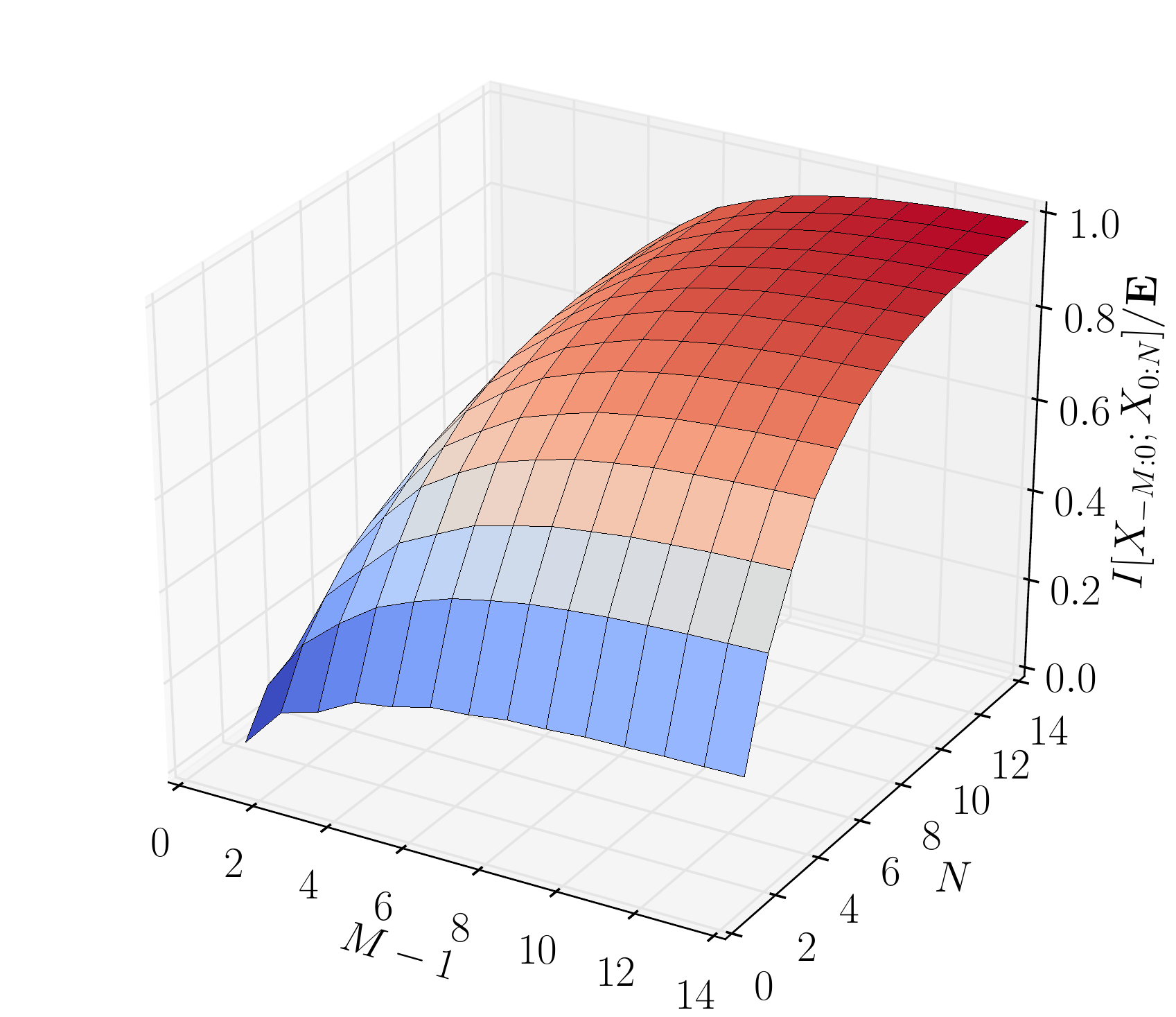}
\caption{Curse of dimensionality when predicting the Even Process:
  (top) The \eM, its minimal unifilar HMM. Edge labels $p|x$ denote generating
  symbol $x \in \MeasAlphabet$ while taking the transition with probability $p$.
  (bottom) $\EE(M,N)/\EE$ as a function of $N$ and $M$ calculated exactly using
  Eq.~(\ref{eq:IpredMN}) and the values of $\{\lambda_i\}$, $\langle
  \delta_{\pi}|W_{\lambda_i} | H(W^{\MeasAlphabet}) \rangle$ from App. I of
  Ref. \cite{Crut13a}. The Even Process's total predictable information
  $\EE \approx 0.9183$ bits. Capturing $90\%$ of $\EE$ requires: $M=6,~N\geq
  13$ or $N=6,~M\geq 13$; $M=7,~N\geq 9$ or $N=7,~M\geq 9$; and $M\geq 8, N\geq
  8$.
  }
\label{fig:EvenProcess_IPredMN}
\end{figure}

Figure \ref{fig:EvenProcess_IPredMN}(bottom) shows $\EE(M,N)$
as a function of $M$ and $N$ for the Even Process, whose \eM\ is displayed in
the top panel. The process's spectral gap is $\approx 0.3$ and,
correspondingly, we see $\EE(M,N)/\EE$ asymptotes slowly to $1$. For example,
capturing $90\%$ of the total predictable information requires $M,N \geq 8$.
(The figure caption contains more detail on allowed $(M,N)$ pairs.) This, in
turn, translates to requiring very good estimates of the probabilities of
$\approx 10^4$ length-$16$ sequences. In Fig. $3$ of Ref. \cite{Stil07b}, by
way of contrast, Even Process information functions were calculated using $M=3$
and $N=2$. As a consequence, the OCF estimates there captured only
$27\%$ of the full $\EE$.

The Even Process is generated by a simple two-state HMM, so it is notable that
computing its information function (done shortly in Sec. \ref{sec:Examples}) is
at all challenging. Then again, the Even Process is an infinite-order Markov
process.

The difficulty can easily become extreme. Altering the Even
Process's lone stochastic transition probability can increase its temporal
correlations such that correctly calculating its information function
requires massive compute resources. Thus, the curse of dimensionality is
a critical concern even for finite-$\Cmu$ processes generated by finite HMMs.

As we move away from such ``simple'' prototype processes and towards real data
sets, the attendant inaccuracies generally worsen. Many natural processes in
physics, biology, neuroscience, finance, and quantitative social science are
highly non-Markovian with slowly asymptoting or divergent $\EE$ \cite{Debo11a}.
This implies rather small spectral gaps if the process has a
countable infinity of causal states---e.g., as in Ref.
\cite{Trav11b}---or a distribution of eigenvalues heavily weighted
near $\lambda=0$, if the process has an uncountable infinity of causal states. In short, ``interesting''
processes \cite{Bial00a} are those for which current information
function algorithms are most likely to fail.

\section{Causal Rate Distortion Theory}
\label{sec:CRD}

Circumventing the curse of dimensionality requires an alternative approach to
predictive rate distortion (PRD)---\emph{causal rate distortion} (CRD)
theory---that leverages the structural information about a process captured by
its forward and reverse causal states. Proposition \ref{prop:CRD}, our first
result, establishes that CRD is equivalent to PRD under certain conditions,
leading to a new and efficient method to calculate information functions for a
broad range of distortion measures. Our second result adapts CRD to
informational distortions, introducing an efficient PIB
alternative---\emph{causal information bottleneck} (CIB). It shows that
compressing $\Past$ to retain information about $\Future$ is equivalent to
compressing $\St^+$ to retain information about $\St^-$. The results generalize
two previous theorems of Refs. \cite{Stil07a, Stil07b, Crut08a, Crut08b}. This
section first describes the relationship between PRD and CRD and then that
between PIB and CIB in an intuitive way. Appendix \ref{sec:App1} gives more
precise statements and their proofs. Sec. \ref{sec:Examples}'s examples
illustrate CRD's benefits, partly by comparing to previous methods and partly
through new analytical insights into information functions.

\begin{figure}[htp]
\centering
\includegraphics[width=0.5\columnwidth]{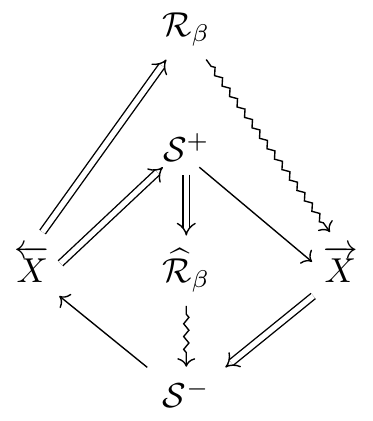} \\
\small{(a) Setting for Causal Rate Distortion Theory.}
\includegraphics[width=0.6\columnwidth]{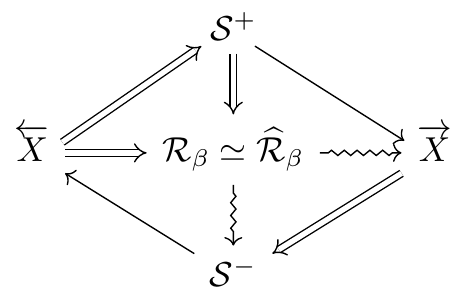}\\
\small{(b) Causal Rate Distortion's Prop. \ref{prop:CRD} and Causal Information
  Bottleneck's Thm. \ref{the:CRD}.}
\caption{Causal Rate Distortion (CRD) Markov chain relationships between the
  past, future, causal states $\FSt$ and $\PSt$, and lossy predictive features
  $\Rep_\beta$ and $\widehat{\Rep}_\beta$.  Diagrammatic notation as in Fig.
  \ref{fig:Intuition2}(bottom).
  }
\label{fig:MarkovChains_CRD}
\end{figure}

To start, inspired by the previous finding that PIB recovers the forward-time
causal states in the zero-temperature ($\beta \to \infty$) limit \cite{Stil07a,
Stil07b}, we argue that compressing either the past $\Past$ or forward-time
causal states $\St^+$ should yield the same predictive features at any
temperature. This leads us to consider two different rate-distortion settings.
In the first, traditional PRD setting we compress the past $\Past$ to minimize
expected distortion about the future $\Future$. In the second, CRD setting we
compress the forward-time causal states $\St^+$ to retain information about the
future $\Future$. Though somewhat intuitive, there is no \emph{a priori} reason
that the two objective functions are related. The Markov chain of Fig.
\ref{fig:MarkovChains_CRD}(top) lays out the implied interdependencies. For many
distortion functions, in fact, they are not related. Lemma \ref{lem:1},
however, says that they are equivalent for certain types of predictive
distortion measures.

{\Lem Compressing the past $\Past$ to minimize expected predictive distortion
is equivalent to compressing the forward-time causal states $\St^+$ to minimize
expected predictive distortion if the distortion measure is expressible as
$d(\Prob(\Future|\Rep=\rep),\Prob(\Future|\Past=\past))$.
\label{lem:1}
}

Many distortion measures can be expressed in this form, though the mapping can
be nonobvious.  For instance, a distortion measure that penalizes the Hamming
distance between the maximum likelihood estimates of the most likely future of
length $L$ is actually expressible in terms of conditional probability
distributions of futures given pasts or features. Lemma \ref{lem:1} then
suggests that PRD will recover the forward-time causal states in the
zero-temperature limit for a great many prediction-related distortion measures,
not just the information distortions considered in Refs. \cite{Stil07a,
Stil07b}.  However, for distortion measures that ask for optimal predictions of
an arbitrary coarse-graining of futures, the forward-time causal states will be
no longer be sufficient statistics. Then, PRD's zero-temperature limit will not
recover the forward-time causal states.

Interestingly, in a nonprediction setting, Ref. \cite{banerjee2004information}
states Lemma~\ref{lem:1} in their Eq.~(2.2) without conditions on the distortion
measure. At least in the context of PRD, this is an oversimplification. For
instance, an entirely reasonable predictive distortion measure could penalize
the difference between the output of a particular prediction algorithm applied
to the true past versus an estimated past. Without proper conditions, these
prediction algorithms can incorporate aspects of the past $\Past$ that are
entirely useless for prediction. A version of Lemma~\ref{lem:1} will still
apply, but the variable that replaces $\Past$ depends on the particular
prediction algorithm. For example, predictions of future sequences of the Even
Process (Sec. \ref{sec:CurseofDimensionality}) based on certain ARIMA estimators will store
unnecessary information about the past. This makes the forward-time causal
states insufficient statistics with respect to the process and the predictor;
see Sec. \ref{sec:retropredictive}. Ideally, prediction algorithms should
tailored to the class of stochastic process to be predicted.

When the distortion measure takes a particular special form, then we can
simplify the objective function further. Our inspiration comes from Refs.
\cite{Crut08a, Crut08b, Crut10d} which showed that the mutual information
between past and future is identical to the mutual information between forward
and reverse-time causal states: $\I[\Past;\Future] = \I[\St^+;\St^-]$. In other
words, forward-time causal states $\St^+$ are the only features needed to
predict the future as well as possible, and reverse-time causal states $\St^-$
are features one \textit{can} predict about the future.

And so, we now consider a third objective function that compresses the
forward-time causal states to minimize expected distortion about the
reverse-time causal states. Proposition \ref{prop:CRD} says that this objective
function is equivalent to compressing the past to minimize expected distortion
for a class of distortion measures.

{\Prop [Causal Rate Distortion] Compressing the past $\Past$ to minimize
expected distortion of the future $\Future$ is equivalent to compressing the
forward-time causal states $\St^+$ to minimize expected distortion of
reverse-time causal states $\St^-$, \emph{if} the distortion measure is
expressible as:
\begin{align*}
d(\past,\rep) = d(\Prob(\Future|\Rep=\rep),\Prob(\Future|\Past=\past))
\end{align*}
and satisfies:
\begin{align*}
d(\Prob(\Future|\Rep & =\rep),\Prob(\Future|\Past=\past)) \\
  & = d(\Prob(\St^-|\Rep=\rep),\Prob(\St^-|\Past=\past))
  ~.
\end{align*}
\label{prop:CRD}
}

Distortion measures that do not satisfy Prop. \ref{prop:CRD}'s conditions, such
as mean squared-error distortion measures, in effect emphasize predicting one
reverse-time causal state over another. Even then, inferring a
maximally-predictive model can improve calculational accuracy for nearly any
distortion measure on sequence distributions. For details, see
App.~\ref{sec:App1}.

Informational distortion measures, though, treat all reverse-time causal states
equally. Leveraging this, Thm.~\ref{the:CRD} follows as a particular
application of Prop.~\ref{prop:CRD}.

{\The [Causal Information Bottleneck] Compressing the past $\Past$ to retain
information about the future $\Future$ is equivalent to compressing $\St^+$ to
retain information about $\St^-$.
\label{the:CRD}
}

Naturally, there is an equivalent version for the time reversed setting in
which past and future are swapped and the causal state sets are swapped. Also,
any forward and reverse-time prescient statistics \cite{Shal98a} can be used in
place of $\St^+$ and $\St^-$ in any of the statements above.

Appendix~\ref{sec:App1}'s proofs follow almost directly from the definitions of
forward- and reverse-time causal states. Variations or portions of Lemma~\ref{lem:1},
Prop.~\ref{prop:CRD}, and Thm.~\ref{the:CRD} may seem intuitive.
Appendix~\ref{sec:App1} points this out when clearly the case.
That said, to the best of our knowledge, Lemma~\ref{lem:1},
Prop.~\ref{prop:CRD}, and
Thm.~\ref{the:CRD} are new.

Theorem~\ref{the:CRD} and Prop.~\ref{prop:CRD} reduce the numerically
intractable problem of clustering in the infinite-dimensional space
$(\Past,\Future)$ to the potentially tractable one of clustering in
$\St^{\pm}$. This is beneficial when a process's causal state set is
finite.  However, many processes have an uncountable infinity of forward-time
causal states or reverse-time causal states \cite{Crut92c, Uppe97a}. Is
Theorem~\ref{the:CRD} useless in these cases? Not at all. A practical
approach is that information functions can be approximated to any desired
accuracy by a finite or countable \eM. However, additional work is required to
understand how model approximations map to information-function approximations.

\section{Examples}
\label{sec:Examples}

To illustrate CRD and CIB, we find lossy causal states and calculate
information functions for processes generated by known \eMs. For several, the
\eMs\ are sufficiently simple that the information functions can be obtained
analytically using the above results. These allow us to comment more generally
on the shape of information functions for several broad classes of process. The
examples also provide a setting in which to compare CIB information functions
to those from \emph{optimal causal filtering} (OCF) \cite{Stil07a, Stil07b}.
Though, Thm.~\ref{the:CRD} says that OCF and CIB yield identical results in the
$L\rightarrow\infty$ limit, the examples give a rather sober illustration of
the substantial errors that arise at finite $L$.

We display the results of the PRD analyses in two ways \footnote{These
information functions are closely related to the more typical information
curves seen in Refs. \cite{Stil07a, Stil07b} and elsewhere, as the
informational distortion is the excess entropy less the predictable information
captured.}. The first is an \emph{information function} that graphs the code
rate $I[\Past;\Rep]$ versus the distortion $I[\Past;\Future|\Rep]$. The second
is a \emph{feature curve} of code rate $I[\Past;\Rep]$ versus inverse
temperature $\beta$. We recall that at zero temperature ($\beta \to \infty$)
the code rate $I[\Past;\Rep] = \FutureCmu$ and the forward-time causal states
are recovered: $\Rep \to \FSt$. At infinite temperature ($\beta = 0$) there is
only a single state that provides no shielding and so the information
distortion limits to $I[\Past;\Future|\Rep] = \EE$. As we will see, these
extremes are useful references for monitoring convergence.

We calculate information functions and feature curves following Ref.
\cite{Tish00a}. (So that the development here is self-contained App.
\ref{app:CRDAlgorithm} reviews this approach, but as adapted to our focus on
prediction.) Given $\Prob(\pst,\fst)$, then, one solves for the
$\Prob(\rep|\st^+)$ and $\Prob(\st^+)$ that maximize the CIB objective function
at each $\beta$:
\begin{align}
\mathcal{L}_{\beta} = \I[\Rep;\PSt] - \beta^{-1} \I[\FSt;\Rep]
  ~,
\label{eq:CRDObjective}
\end{align}
by iterating the dynamical system:
\begin{align}
{\Prob}_{t}(\rep|\fst) & = \frac{\Prob_{t-1}(\rep)}{Z_{t}(\fst,\beta)} e^{-\beta
\DKL[\Prob(\pst|\fst) || \Prob_{t-1}(\pst|\rep)]} \label{eq:CRD1} \\
  {\Prob}_t(\rep) & = \sum_{\fst} \Prob_t(\rep|\fst) \Prob(\fst) \\
  {\Prob}_t(\pst|\rep) & = \sum_{\fst} \Prob(\pst|\fst) {\Prob}_t(\fst|\rep)
  ~,
\label{eq:CRD3}
\end{align}
where $Z_t(\fst,\beta)$ is the normalization constant for
$\Prob_{t}(\rep|\fst)$. Iterating Eqs.~(\ref{eq:CRD1}) and (\ref{eq:CRD3})
gives (i) one point on the function $(R_{\beta},D_{\beta})$ and (ii) the
explicit optimal lossy predictive features $\Rep_\beta =
\{\Prob_t (\rep|\fst)\}$.

For each $\beta$, we chose $500$ random initial $\Prob_0(\rep|\fst)$, iterated
Eqs. (\ref{eq:CRD1})-(\ref{eq:CRD3}) $300$ times, and recorded the solution
with the largest $\mathcal{L}_{\beta}$. This procedure finds local maxima of
$\mathcal{L}_{\beta}$, but does not necessarily find global maxima. Thus, if
the resulting information function was nonmonotonic, we increased the number of
randomly chosen initial $\Prob_0(\rep|\fst)$ to $5000$, increased the number of
iterations to $500$, and repeated the calculations. This brute force approach
to the nonconvexity of the objective function was feasible here only due to
analyzing processes with small \eMs. Even so, the estimates might include
suboptimal solutions in the lossier regime. A more sophisticated approach would
leverage the results of Ref. \cite{parker2002annealing, parker2004bifurcation,
parker2010symmetry} to move carefully from high-$\beta$ to low-$\beta$
solutions.

We used a similar procedure to calculate OCF functions, but $\fst$ and $\pst$
were replaced by $\ms_{-L:0}$ and $\ms_{0:L}$, which were then replaced by
finite-time causal states $\St^+_{L,L}$ and $\St^-_{L,L}$ using a finite-time
variant of CIB. The joint probability distribution of these finite-time causal
states was calculated exactly by (i) calculating sequence distributions of
length $2L$ directly from the \eM\ transition matrices and (ii) clustering
these into finite-time causal states using the equivalence relation described
in Sec. \ref{sec:Background_CM}, except when the joint probability
distribution was already analytically available. This procedure avoids the
complications of finite data samples. As a result, differences between the
results produced by CIB and OCF are entirely a difference in the objective
function.

Note that in contrast with deterministic annealing procedures that start at low
$\beta$ (high temperature) and add codewords to expand the codebook as
necessary, CRD algorithms can start at large $\beta$ with a codebook with
codewords $\St^+$ and decrease $\beta$, allowing the representation to
naturally reduce its size.  This is usually ``naive''
\cite{Elidan:2002:IBE:2100584.2100608} due to the large number of local maxima
of $\mathcal{L}_{\beta}$, but here, we know the zero-temperature result
beforehand.  Of course, CRD algorithms could also start at low $\beta$ and
increase $\beta$.  The key difference between CRD and PRD, and between CIB and
PIB, is not the algorithm itself, but the joint probability distribution of
compressed and relevant variables.  

Section \ref{sec:elusive_info} gives conditions on a process which guarantee
that its information functions can be accurately calculated \textit{without}
first having a maximally-predictive model in hand. Section
\ref{sec:retropredictive} describes several processes that have first-order
phase transitions in their feature curves at $\beta=1$. Section
\ref{sec:RIP_example} describes how information functions and feature curves
can change nontrivially under time reversal. Finally, Sec. \ref{sec:tent_map}
shows how predictive features describe predictive ``macrostates'' for the
process generated the symbolic dynamics of the chaotic Tent Map.

\subsection{Unhidden and Almost Unhidden Processes}
\label{sec:elusive_info}

PIB algorithms that cluster pasts of length $M \geq 1$ to retain information
about futures of length $N\geq 1$ calculate accurate information functions when
$\EE(M,N) \approx \EE$. (Recall Sec. \ref{sec:CurseofDimensionality}.) Such
algorithms work exactly on order-$R$ Markov processes when $M,N\geq R$, since
$\EE(R,R) = \EE$. However, there are many processes that are ``almost''
order-$R$ Markov, for which algorithms based on finite-length pasts and futures
should work quite well.

The inequality of Eq. (\ref{eq:OCF_Bound}) suggests that, as far as accuracy is concerned, if a process has a small $\sigmamu(L)$ relative to its $\EE$ for some reasonably small $L$, then sequences are effective states. This translates into the conclusion that for this class of process calculating information functions by first moving to causal state space is unnecessary.

\begin{figure}[htp]
\centering
\includegraphics[width=0.7\columnwidth]{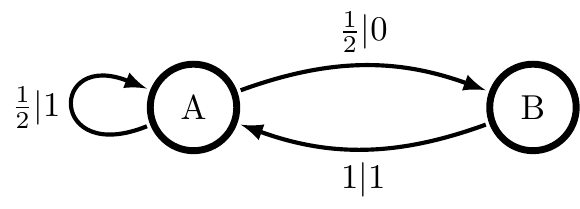}
\includegraphics[width=0.9\columnwidth]{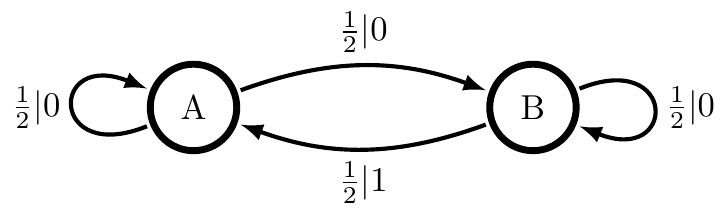}
\caption{(top) Golden Mean HMM, an \eM. (bottom) Simple Nonunifilar Source
  nonunifilar HMM presentation; not the SNS process's \eM.
  }
\label{fig:SNS_HMM}
\end{figure}

\begin{figure}[htp]
\centering
\includegraphics[width=\columnwidth]{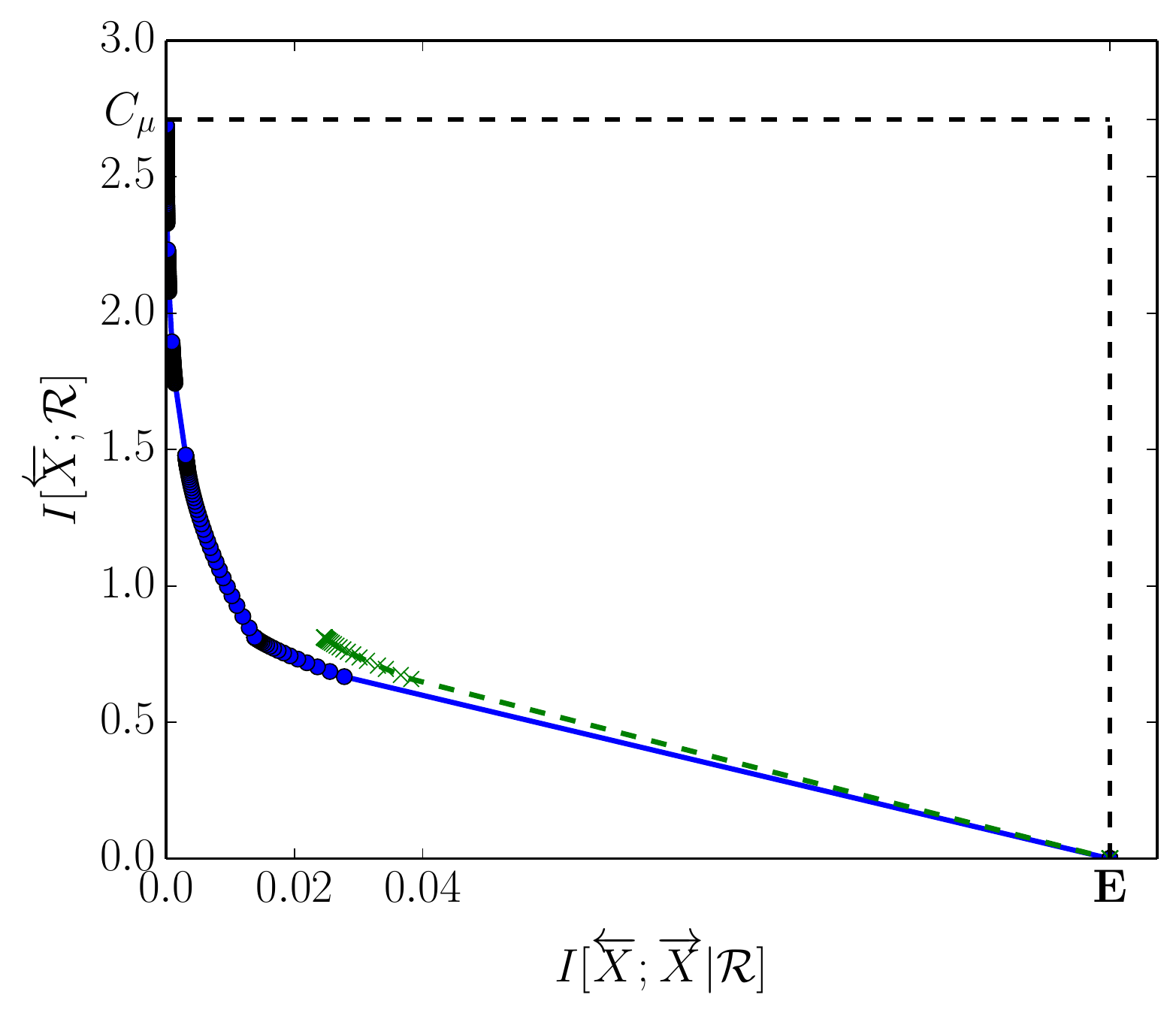}
\includegraphics[width=\columnwidth]{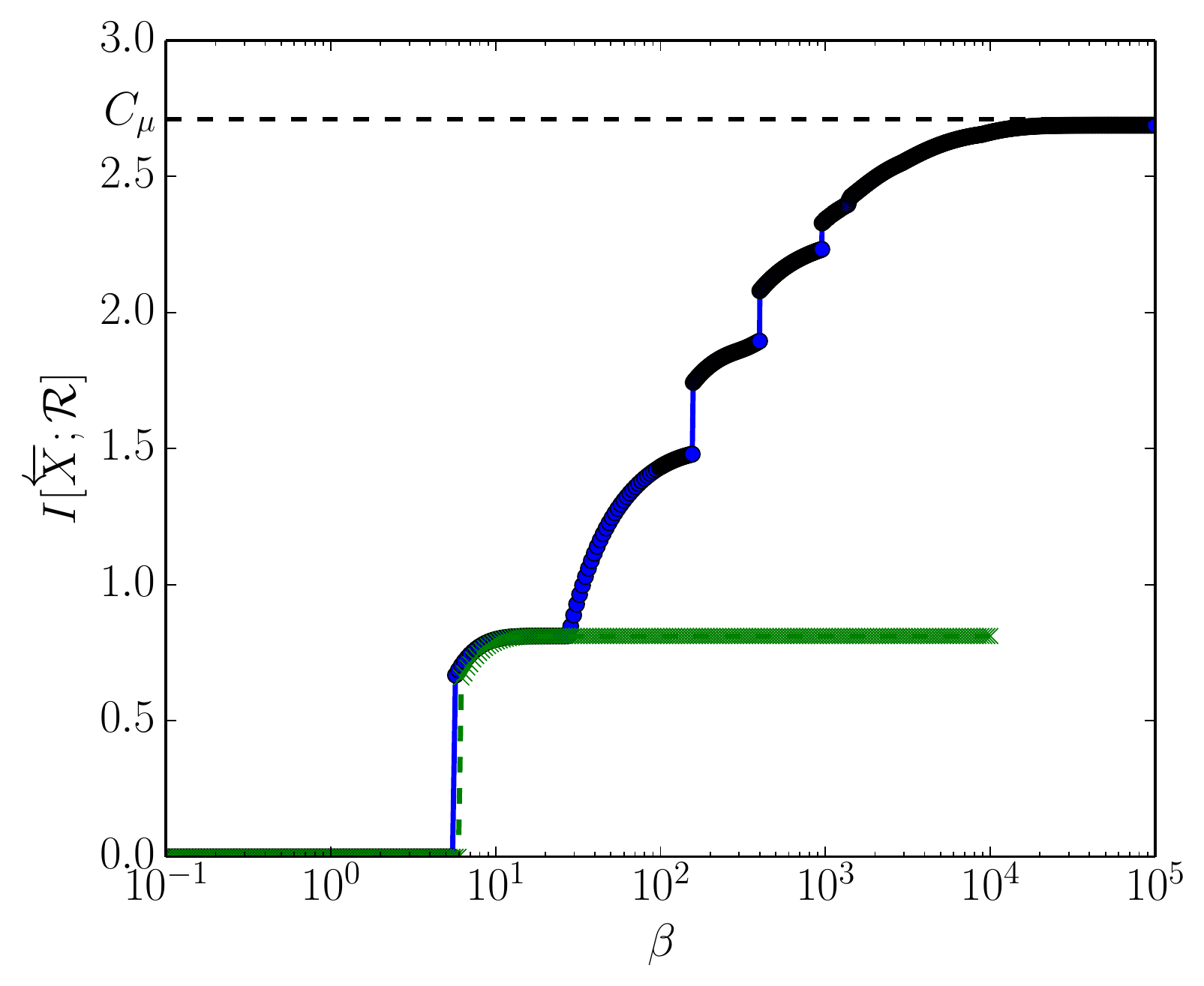}
\caption{Simple Nonunifilar Source:
  (Top panel) Information function: coding cost versus distortion.
  (Bottom panel) Feature curve: coding cost as a function of inverse
  temperature $\beta$.
  (Blue solid line, circles) CIB with a $10$-state approximate \eM.
  (Green dashed line, crosses) OCF at $M,N = 1$.
  }
\label{fig:SNS_PIB}
\end{figure}

Let's test this intuition. The prototypical example with $\sigmamu(1) =0$ is the Golden Mean Process, whose HMM is shown in Fig.~\ref{fig:SNS_HMM}(top). It is order-$1$ Markov, so OCF with $L=1$ is provably equivalent to CIB, illustrating one side of the intuition.

A more discerning test is an infinite-order Markov process with small
$\sigmamu$. One such process is the Simple Nonunifilar Source (SNS) whose
(nonunifilar) HMM is shown in Fig.~\ref{fig:SNS_HMM}(bottom). As anticipated,
Fig.~\ref{fig:SNS_PIB}(top) shows that OCF with $L=1$ and CIB yield very similar
information functions at low code rate and low $\beta$. In fact, many of SNS's
statistics are well approximated by the Golden Mean HMM.

The feature curve in Fig.~\ref{fig:SNS_PIB}(bottom) reveals a slightly more
nuanced story, however. The SNS is highly cryptic, in that it has a much larger
$\Cmu$ than $\EE$. As a result, OCF with $L=1$ approximates $\EE$ quite well
but underestimates $\Cmu$, replacing an (infinite) number of feature-discover
transitions with a single transition. (More on these transitions shortly.i)

This particular type of error---missing predictive features---only matters for
predicting the SNS when low distortion is desired. Nonetheless, it is important
to remember that the process implied by OCF with $L=1$---the Golden Mean Process---is not the SNS. The Golden Mean Process is an order-$1$ Markov process. The SNS HMM is nonunifilar and generates an infinite-order Markov process and so provides a classic example \cite{Crut92c} of how difficult it can be to exactly calculate information measures of stochastic processes.

Be aware that CIB cannot be directly applied to analyze the SNS, since the
latter's causal state space is countably infinite; see Ref. \cite{Marz14b}'s
Fig. 3. Instead, we used finite-time causal states with finite past and future
lengths and with the state probability distribution given in App. B of Ref.
\cite{Marz14b}. Here, we used $M,N = 10$, effectively approximating the SNS as
an order-$10$ Markov process.

\subsection{First-order Phase Transitions at $\beta=1$}
\label{sec:retropredictive}

Feature curves have discontinuous jumps (``first-order phase transitions'') or
are nondifferentiable (``second-order phase transitions'') at critical
temperatures when new features or new lossy causal states are discovered.  The
effective dimension of the codebook changes at these transitions.  Symmetry
breaking plays a key role in identifying the type and temperature of phase
transitions in constrained optimization \cite{Rose94a,parker2010symmetry}.
Using the infinite-order Markov Even Process of Sec.
\ref{sec:CurseofDimensionality}, CIB allows us to explore in greater detail why
and when first-order phase transitions occur at $\beta=1$ in feature curves.

\begin{figure}[htp]
\centering
\includegraphics[width=\columnwidth]{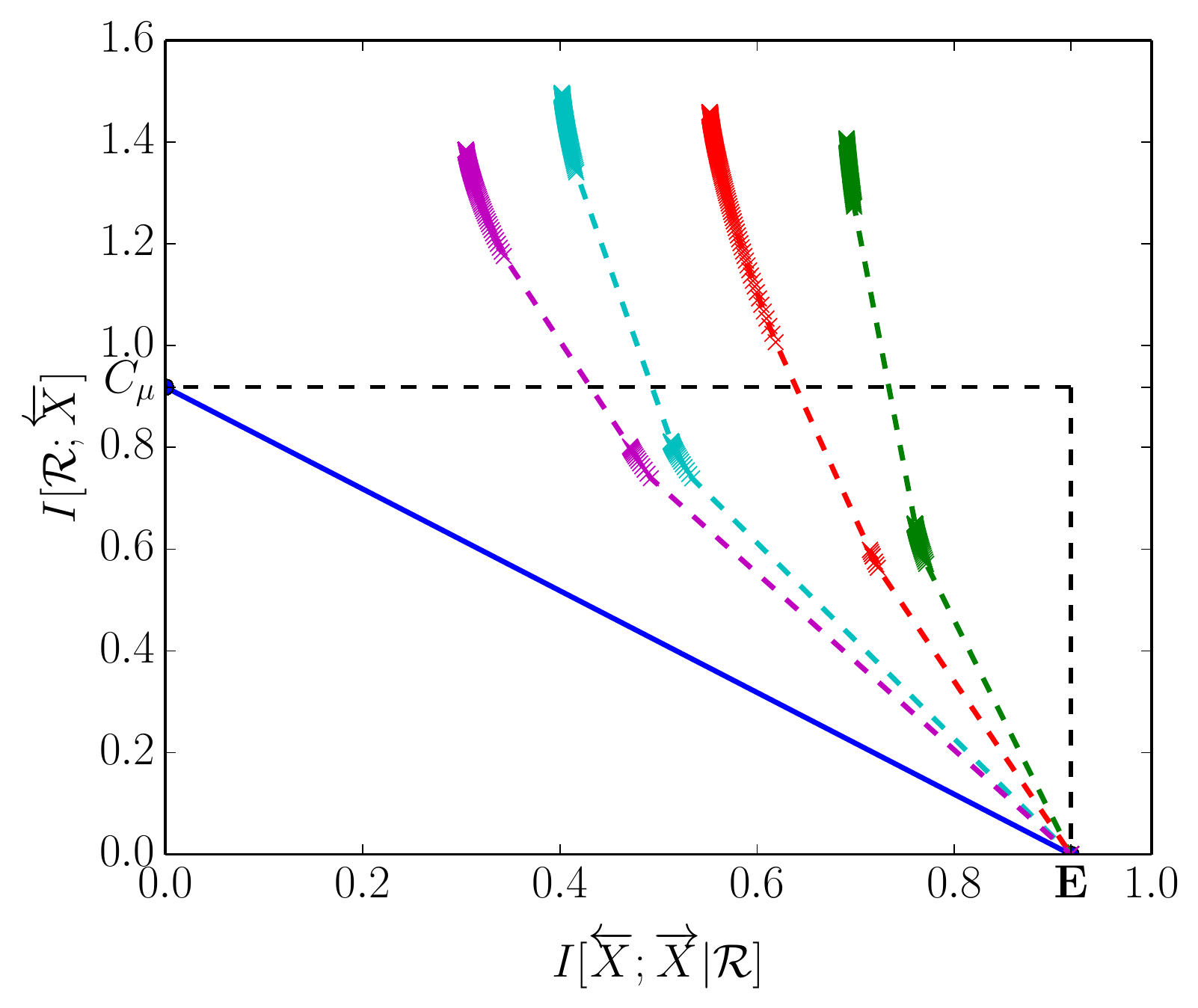}
\includegraphics[width=\columnwidth]{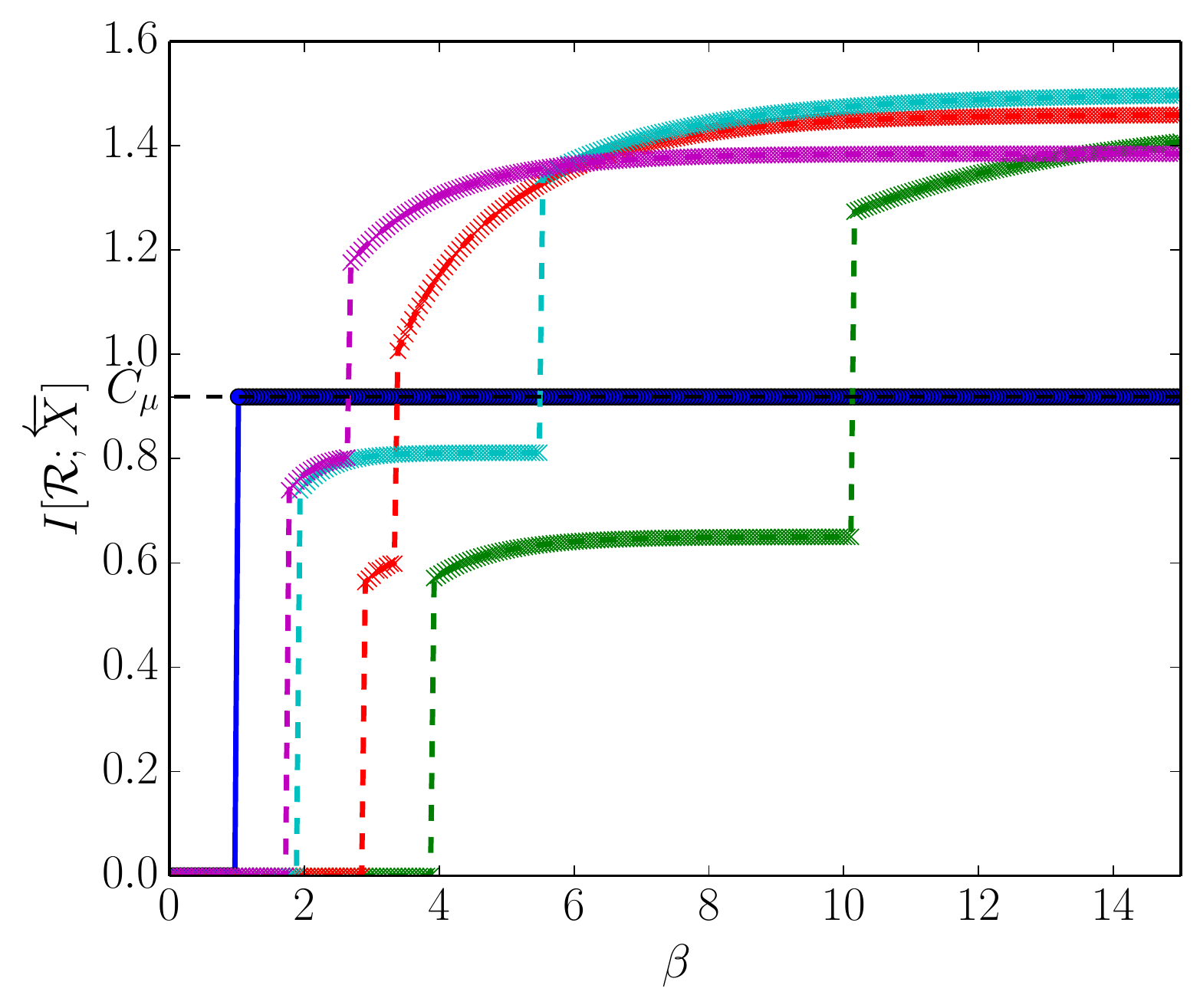}
\caption{Even Process analyzed with CIB (solid line, blue circles) and with OCF
  (dashed lines, colored crosses) at various values of $M = N = L$: (right to
  left) $L=2$ (green), $L=3$ (red), $L=4$ (light blue), and $L=5$ (purple).
  (top) Information functions. (bottom) Feature curves.
  At $\beta=1$, CIB functions transition from approximating the Even Process as
  i.i.d. (biased coin flip) to identifying both causal states.
  }
\label{fig:EvenProcess_PIB}
\end{figure}

There are important qualitative differences between information functions and
feature curves obtained via CIB and via OCF for the Even Process. First, as
Fig.~\ref{fig:EvenProcess_PIB}(top) shows, the Even Process CIB information
function is a simple straight line, whereas those obtained from OCF are curved
and substantially overestimate the code rate.  Second, as
Fig.~\ref{fig:EvenProcess_PIB}(bottom) shows, the CIB feature curve is
discontinuous at $\beta=1$, indicating a single first-order phase transition
and the discovery of highly predictive states. In contrast, OCF functions miss
that key transition and incorrectly suggest several phase transitions at larger
$\beta$s.

The first result is notable, as Ref. \cite{Stil07b} proposed that the curvature
of OCF information functions define natural scales of predictive
coarse-graining. In this interpretation, linear information functions imply
that the Even Process has \textit{no} such intermediate natural scales. And,
there are good reasons for this.

So, why does the Even Process exhibit a straight line? Recall that the Even
Process's recurrent forward-time causal states code for whether or not one just
saw an even number of $1$'s (state A) or an odd number of $1$'s (state B) since
the last $0$. Its recurrent reverse-time causal states (Fig. 2 in Ref.
\cite{Crut08b}) capture whether or not one will see an even number of $1$'s
until the next $0$ or an odd number of $1$'s until the next $0$. Since one only
sees an even number of $1$'s between successive $0$'s, knowing the forward-time
causal state uniquely determines the reverse-time causal state and vice versa.
The Even Process' forward causal-state distribution is $\Prob(\St^+) = \big(
2/3 ~ 1/3 \big)$  and the conditional distribution of forward and reverse-time
causal states is:
\begin{align*}
\Prob(\St^-|\St^+) = \begin{pmatrix} 1 & 0 \\ 0 & 1 \end{pmatrix}
  ~.
\end{align*}
Thus, there is an invertible transform between $\St^+$ and $\St^-$, a
conclusion that follows directly from the process's bidirectional machine.
The result is that:
\begin{align}
\I[\Rep;\St^+] = \I[\Rep;\St^-]
  ~.
\label{eq:EvenProcess_Stpm}
\end{align}
And so, we directly calculate the information function from Eq.
(\ref{eq:InfoCurve}):
\begin{align*}
R(I_0) &= \min_{\I[\Rep;\St^-] \geq I_0} \I[\Rep;\St^+] \\
  & = \min_{\I[\Rep;\St^-] \geq I_0} \I[\Rep;\St^-] \\
  & = I_0
  ~,
\end{align*}
for all $I_0\leq \EE$. Similar arguments hold for periodic process as described
in Ref. \cite{Stil07a, Stil07b} and for general \emph{cyclic} (noisy periodic)
processes as well. However, periodic processes are finite-order Markov, whereas
the infinite Markov-order Even Process hides its deterministic relationship between prediction and
retrodiction underneath a layer of stochasticity. This suggests that the
bidirectional machine's \emph{switching maps} \cite{Crut08b} are key to the
shape of information functions.

The Even Process's feature curve in (Fig.~\ref{fig:EvenProcess_PIB}(bottom) shows a
first-order phase transition at $\beta = 1$.  Similar to periodic and cyclic
processes, its lossy causal states are all-or-nothing. Iterating Eqs. (\ref{eq:CRD1}) and (\ref{eq:CRD3}) is an attempt to maximize
the objective function of Eq. (\ref{eq:CRDObjective}). However,
Eq.~(\ref{eq:EvenProcess_Stpm}) gives:
\begin{align*}
\mathcal{L}_{\beta} = (1-\beta^{-1}) \I[\Rep;\St^+]
  ~.
\end{align*}
Recall that $0\leq \I[\Rep;\St^+]\leq \Cmu$. For $\beta<1$, on the one hand,
maximizing $\mathcal{L}_{\beta}$ requires minimizing $\I[\Rep;\St^+]$, so the
optimal lossy model is an i.i.d. approximation of the Even Process---a
single-state HMM. For $\beta>1$, on the other, maximizing $\mathcal{L}_{\beta}$
requires maximizing $\I[\Rep;\St^+]$, so the optimal lossy features are the
causal states $A$ and $B$ themselves. At $\beta=1$, though,
$\mathcal{L}_{\beta}=0$, and any representation $\Rep$ of the forward-time
causal states $\St^+$ is optimal. In sum, the discontinuity of coding cost
$\I[\Rep;\St^+]$ as a function of $\beta$ corresponds to a first-order phase
transition and the critical inverse temperature is $\beta=1$.

Both causal states in the Even Process are unusually predictive features: any increase in memory of such causal states is accompanied by a proportionate increase in predictive power. These states are associated with a one-to-one (switching) map between a forward-time and reverse-time causal state. In principle, such states should be the first features extracted by any PRD algorithm. More generally, when the joint probability distribution of forward- and reverse-time causal states can be permuted into diagonal block-matrix form, there should be a first-order phase transition at $\beta=1$ with one new codeword for each of the blocks.

Many processes do not have probability distributions over causal states that
can be permuted, even approximately, into a diagonal block-matrix form; e.g.,
most of those described in Refs. \cite{Marz14a, Marz14b}. However, we suspect
that diagonal block-matrix forms for $\Prob(\St^+,\St^-)$ might be relatively
common in the highly structured processes generated by low-entropy rate
deterministic chaos, as such systems often have many irreducible forbidden
words. Restrictions on the support of the sequence distribution easily yields
blocks in the joint probability distribution of forward- and reverse-time
causal states.

For example, the Even Process forbids words with an odd number of $1$s,
which is expressed by its \emph{irreducible forbidden word} list
$\mathcal{F} = \{0 1^{2k+1}0: k = 0, 1, 2, \ldots \}$. Its causal states group
pasts that end with an even (state $A$) or odd (state $B$) number of $1$s since
the last $0$. Given the Even Process' forbidden words $\mathcal{F}$, sequences
following from state $A$ must start with an even number of ones before the next
$0$ and those from state $B$ must start with an odd number of ones before the
next $0$. The restricted support of the Even Process' sequence distribution
therefore gives its causal states substantial predictive power.

Moreover, many natural processes are produced by deterministic chaotic maps
with added noise \cite{Crut82a}. Such processes may also have
$\Prob(\St^+,\St^-)$ in \textit{nearly} diagonal block-matrix form. These
joint probability distributions might be associated with sharp second-order
phase transitions.

However, numerical results for the ``four-blob'' problem studied in Ref.
\cite{parker2010symmetry} suggest the contrary. The joint probability
distribution of compressed and relevant variables there has a nearly diagonal
block-matrix form, with each block corresponding to one of the four blobs. If
the joint probability distribution were exactly block diagonal---e.g., from a
truncated mixture of Gaussians model---then the information function would be
linear and the feature curve would exhibit a single first-order phase
transition at $\beta=1$ from the above arguments. The information function for
the four-blob problem looks linear; see Fig. $5$ of Ref.
\cite{parker2010symmetry}.  The feature curve (Fig. $4$, there) is entirely
different from the feature curves that we would expect from our earlier
analysis of the Even Process. Differences in the off-diagonal block-matrix
structure allowed the annealing algorithm to discriminate between the nearly
equivalent matrix blocks, so that there are three phase transitions to identify
each of the four blobs. Moreover, none of the phase transitions are sharp. So,
perhaps the sharpness of phase transitions in feature curves of noisy chaotic
maps might have a singular noiseless limit, as is often true for information
measures \cite{Marz14a}.

\begin{figure}[htp]
\centering
\includegraphics[width=0.75\columnwidth]{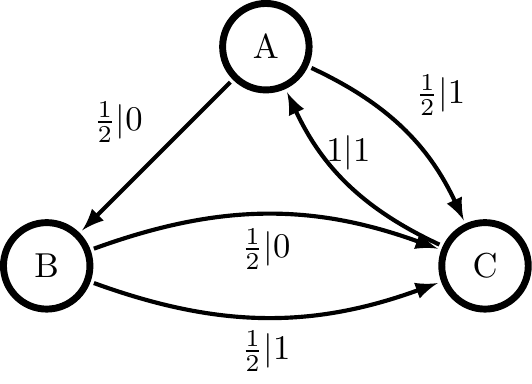}
\includegraphics[width=0.8\columnwidth]{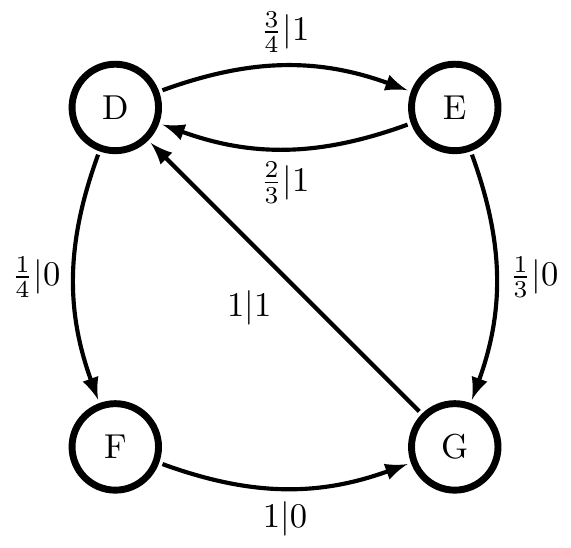}
\caption{Random Insertion Process (RIP):
  (Top) Forward-time \eM. (Bottom) Reverse-time \eM.
  }
\label{fig:RIP_eM}
\end{figure}

\begin{figure}[htp]
\centering
\includegraphics[width=\columnwidth]{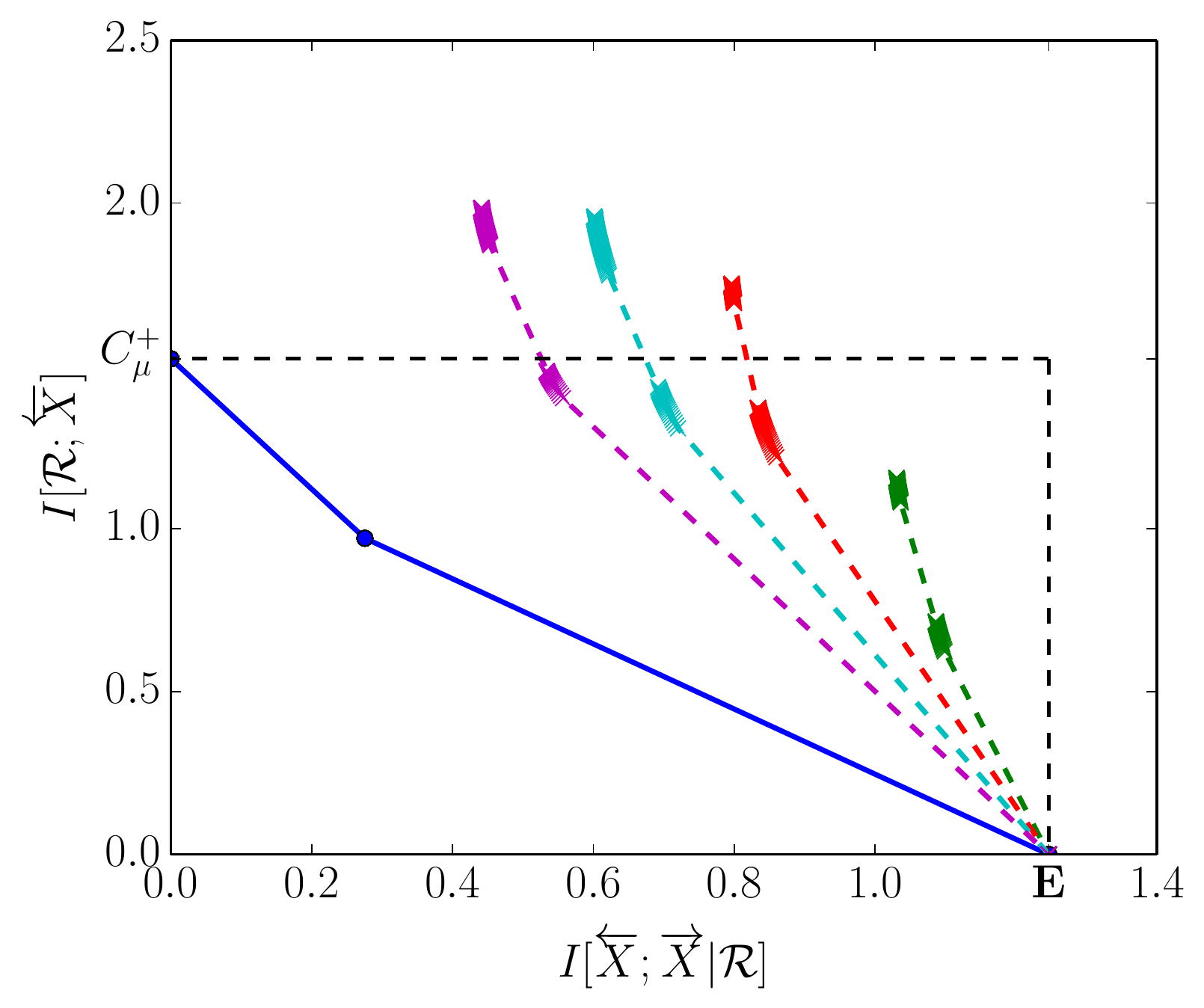}
\includegraphics[width=\columnwidth]{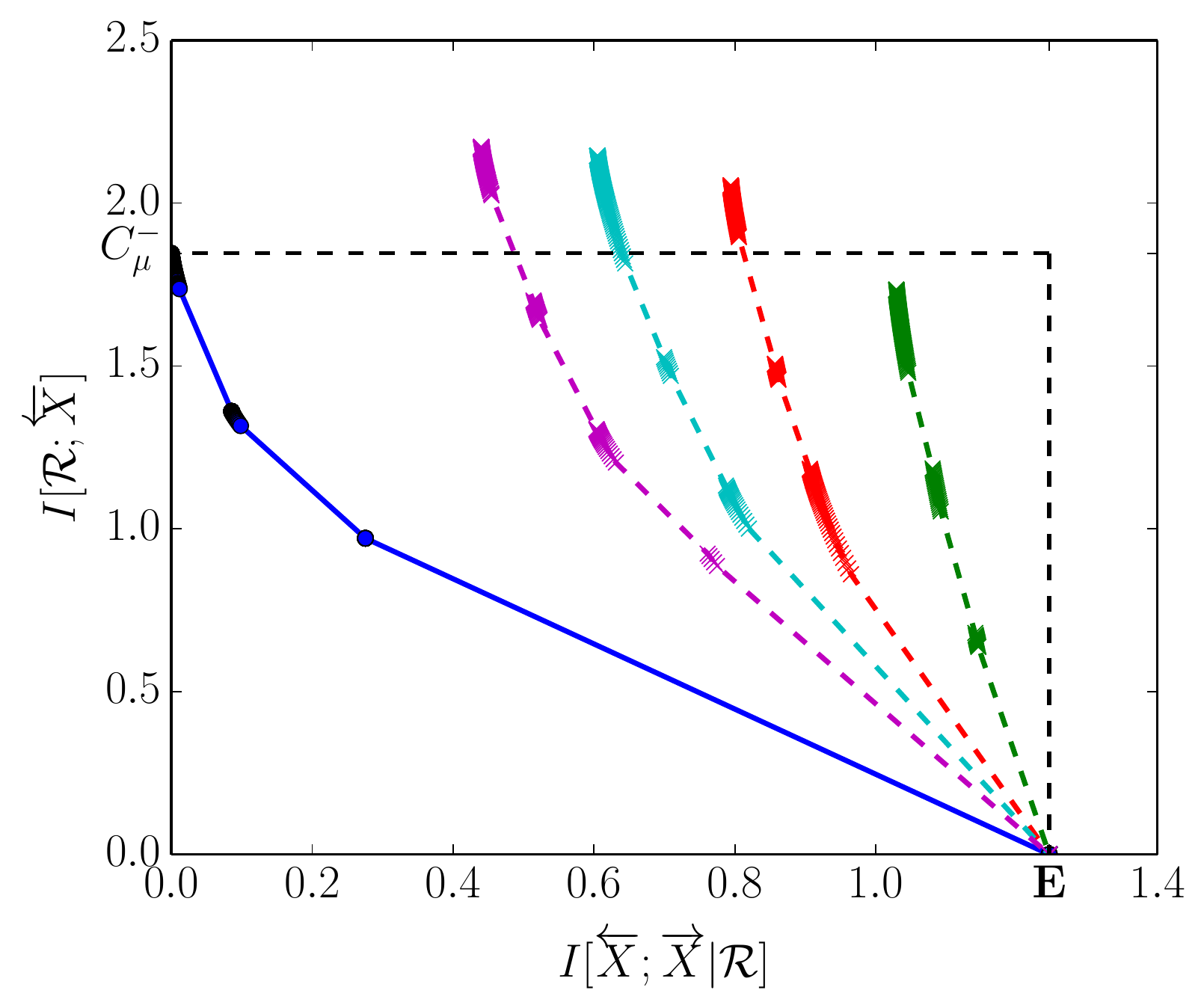}
\caption{Random Insertion Process (RIP) Information Functions:
  RIP is a causally irreversible process: $\FutureCmu < \PastCmu$.  There are
  more causal states in reverse time than forward time, leading to more kinks
  in the reverse-time process' information function (bottom) than in the
  forward-time process' information function (top). Legend as in previous
  figure: (solid line, blue circles) CIB function and (dashed lines, colored crosses)
  OCF at various sequence lengths.
  }
\label{fig:RIP_TransitionOrder1}
\end{figure}

\begin{figure}[htp]
\centering
\includegraphics[width=\columnwidth]{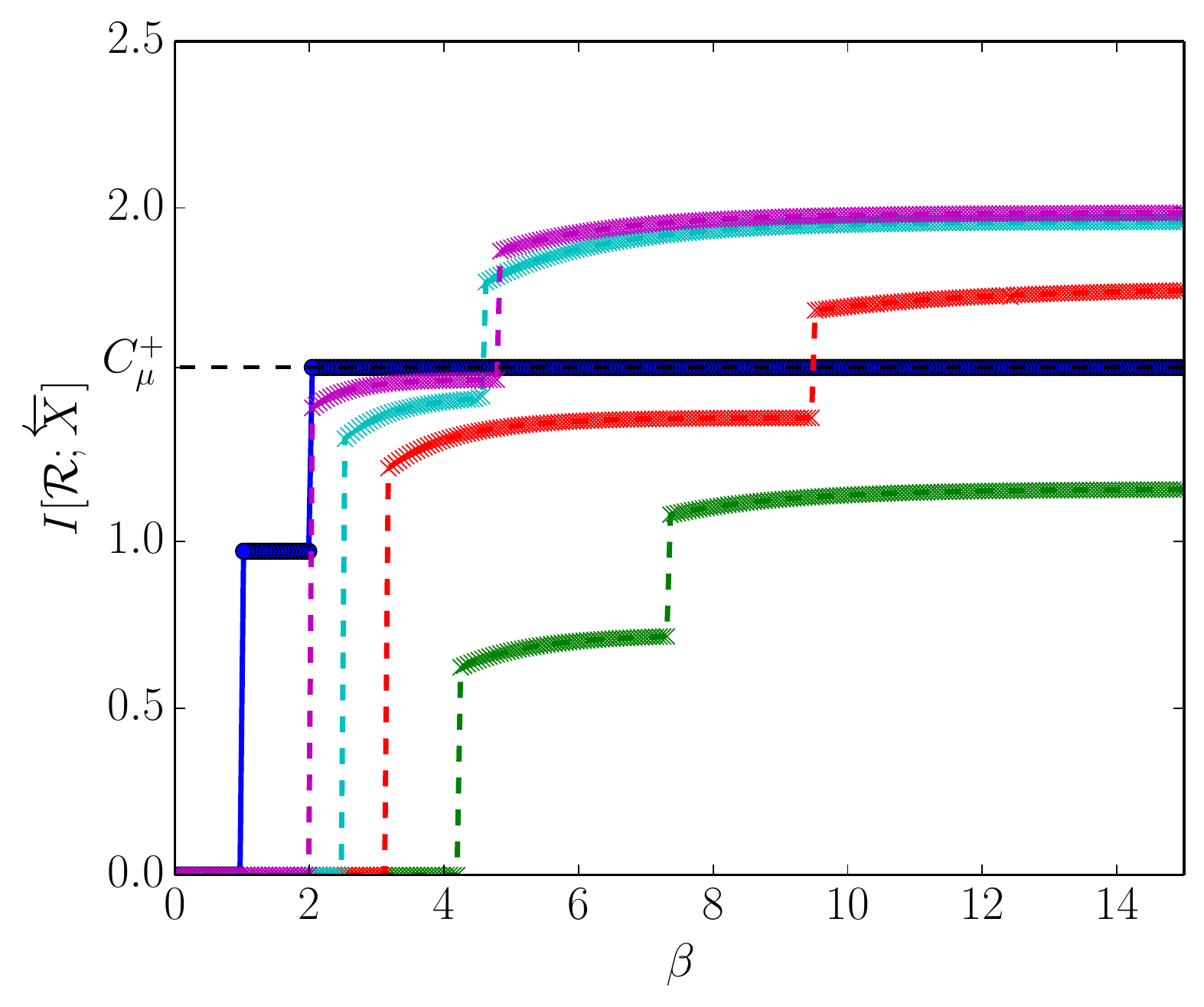}
\includegraphics[width=\columnwidth]{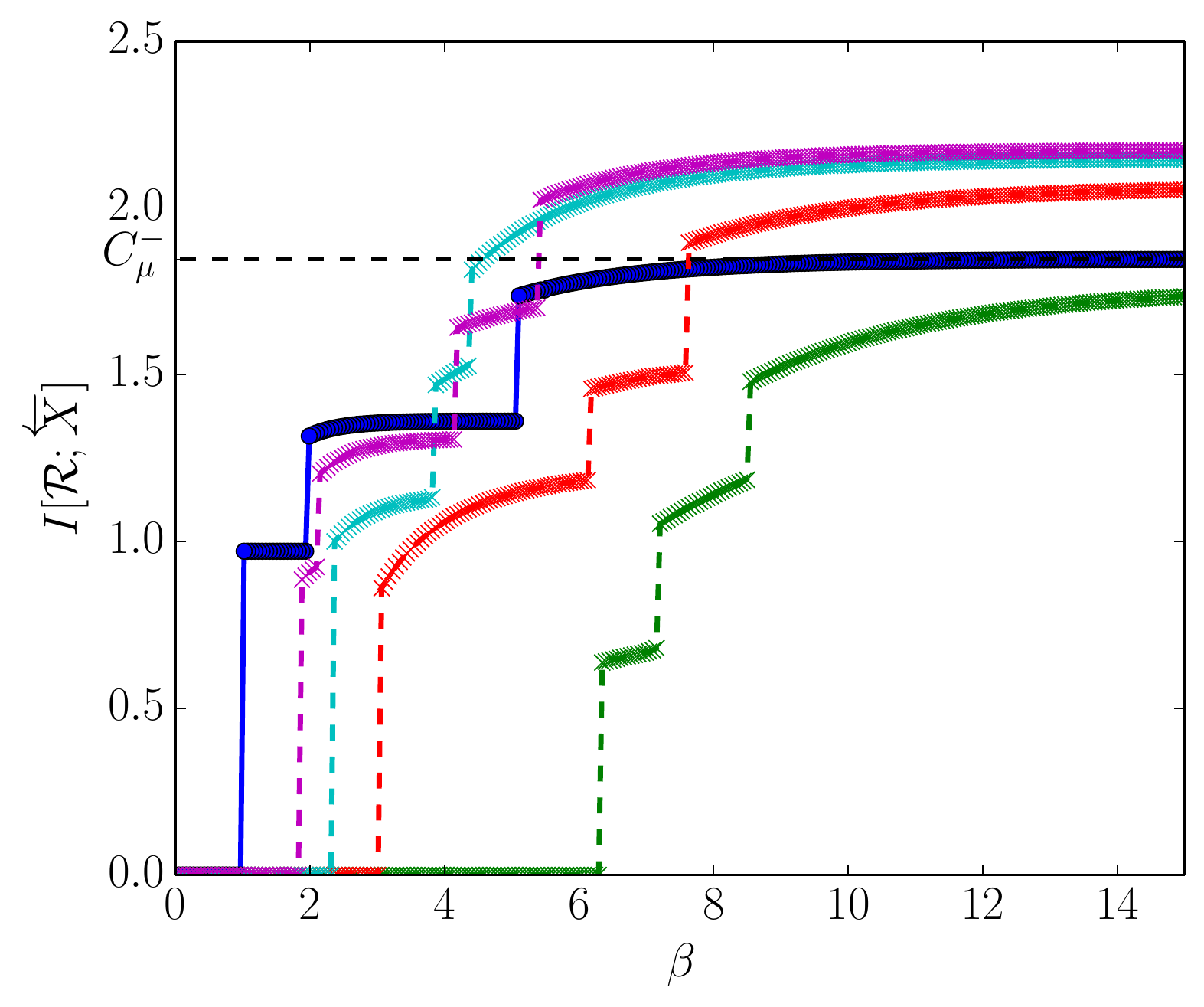}
\caption{Random Insertion Process (RIP) Feature Curves: Having more causal
  states in reverse time than forward time leads to more phase transitions in
  the reverse-time process' feature curve (bottom) than in the forward-time
  process' feature curve (top). Legend as in previous figure.
  }
\label{fig:RIP_TransitionOrder}
\end{figure}

\subsection{Temporal Asymmetry in Lossy Prediction}
\label{sec:RIP_example}

As Refs. \cite{Crut08a, Crut08b} describe, the resources required to losslessly
predict a process can change markedly under time reversal. The prototype example is the Random Insertion Process (RIP), shown in Fig.
~\ref{fig:RIP_eM}. Its bidirectional machine is known analytically
\cite{Crut08a}. Therefore, we know the joint $\Prob(\FSt,\PSt)$ via
$\Prob(\St^+) = \big( 2/5 ~ 1/5 ~ 2/5 \big)$
and:
\begin{align*}
   \Prob(\St^-|\St^+) =
   \begin{pmatrix} 0 & \frac{1}{2} & 0 & \frac{1}{2} \\
   0 & \frac{1}{2} & \frac{1}{2} & 0 \\
   1 & 0 & 0 & 0 \end{pmatrix}
  ~.
\end{align*}
There are three forward-time causal states and four reverse-time causal states,
and the forward-time statistical complexity and reverse-time statistical
complexity are unequal, making the RIP causally irreversible. For instance,
$\FutureCmu \approx 1.8$ bits and $\PastCmu \approx 1.5$ bits, even though the
excess entropy $\EE\approx 1.24$ bits is by definition time-reversal invariant.

However, it could be that the lossy causal states are somehow more robust to
time reversal than the (lossless) causal states themselves. Let's investigate
the difference in RIP's information and feature curves under time reversal.
Figure~\ref{fig:RIP_TransitionOrder1} shows information functions for the
forward-time and reverse-time processes. Despite RIP's causal irreversibility,
information functions look similar until informational distortions of less than
$0.1$ bits. RIP's temporal correlations are sufficiently long-ranged so as to
put OCF with $L\leq 5$ at a significant disadvantage relative to CIB, as the
differences in the information functions demonstrate. OCF greatly
underestimates $\EE$ by about $30\%$ and both underestimates and overestimates
the correct $\Cmu$.

The RIP feature curves in Fig.~\ref{fig:RIP_TransitionOrder} reveal a similar
story in that OCF fails to asymptote to the correct $\Cmu$ for any $L\leq 5$ in
either forward or reverse time. Unlike the information functions, though,
feature curves reveal temporal asymmetry in the RIP even in the lossy (low
$\beta$) regime.

Both forward and reverse-time feature curves show a first-order phase
transition at $\beta=1$, at which point the forward-time causal state $C$ and
the reverse-time causal state $D$ are added to the codebook, illustrating the
argument of Sec. \ref{sec:retropredictive}. (Forward-time causal state $C$ and
reverse-time causal state $D$ are equivalent to the same bidirectional causal
state $C/D$ in RIP's bidirectional \eM. See Fig. 2 of Ref. \cite{Crut08a}.)
This common bidirectional causal state is the main source of similarity in the
information functions of Fig.~\ref{fig:RIP_TransitionOrder1}.

Both feature curves also show phase transitions at $\beta=2$, but similarities
end there. The forward-time feature curve shows a first-order phase transition
at $\beta=2$, at which point both remaining forward-time causal states $A$ and
$B$ are added to the codebook. The reverse-time feature curve has what looks to
be a sharp second-order phase transition at $\beta=2$, at which point the
reverse-time causal state $F$ is added to the codebook. The remaining two
reverse-time causal states, $E$ and $G$, are finally added to the codebook at
$\beta=5$. We leave solving for the critical temperatures and confirming the
phase transition order using a bifurcation discriminator
\cite{parker2002annealing} to the future.

\subsection{Predictive Hierarchy in a Dynamical System}
\label{sec:tent_map}

\begin{figure}[htp]
\centering
\includegraphics[width=\columnwidth]{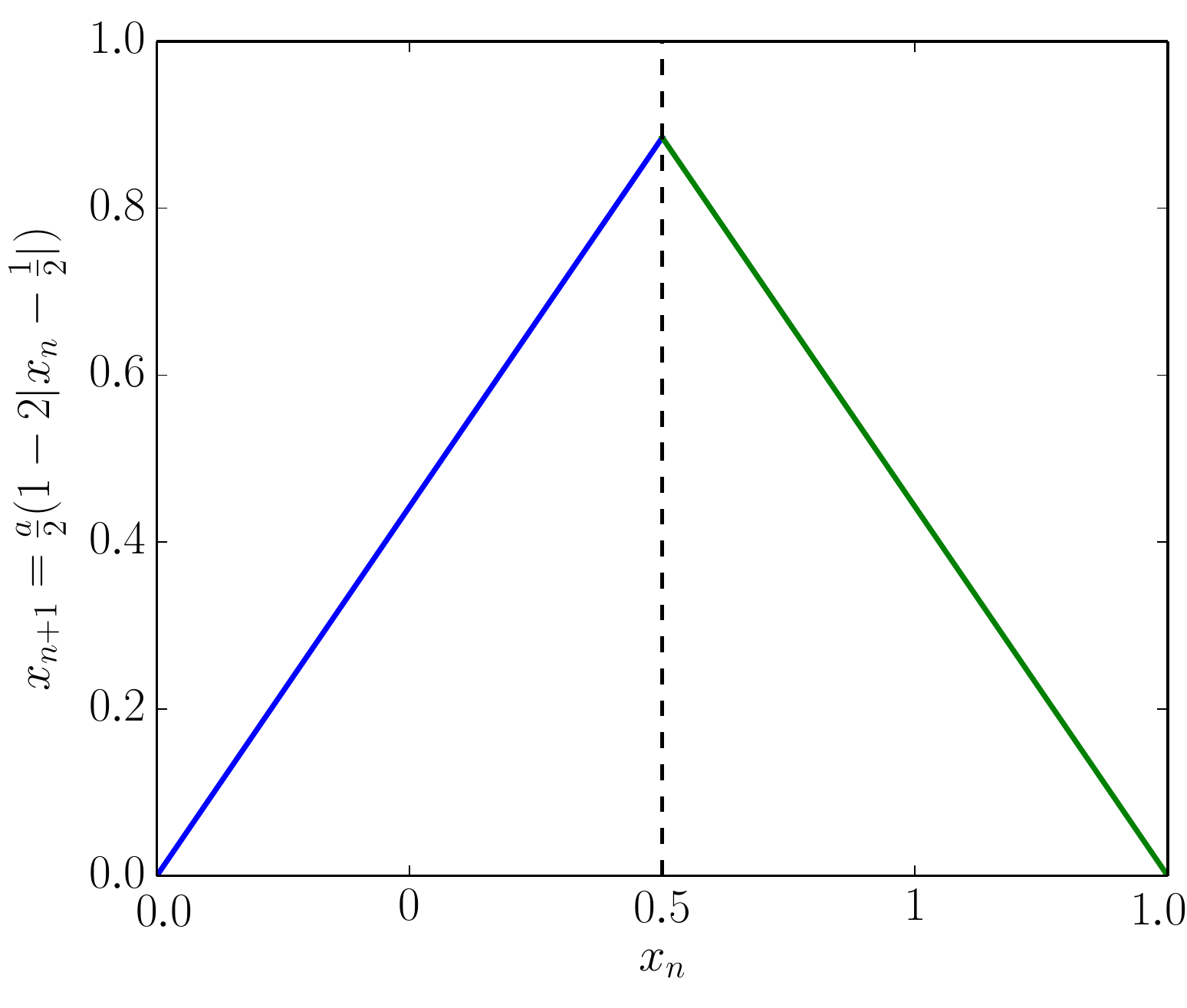}
\includegraphics[width=\columnwidth]{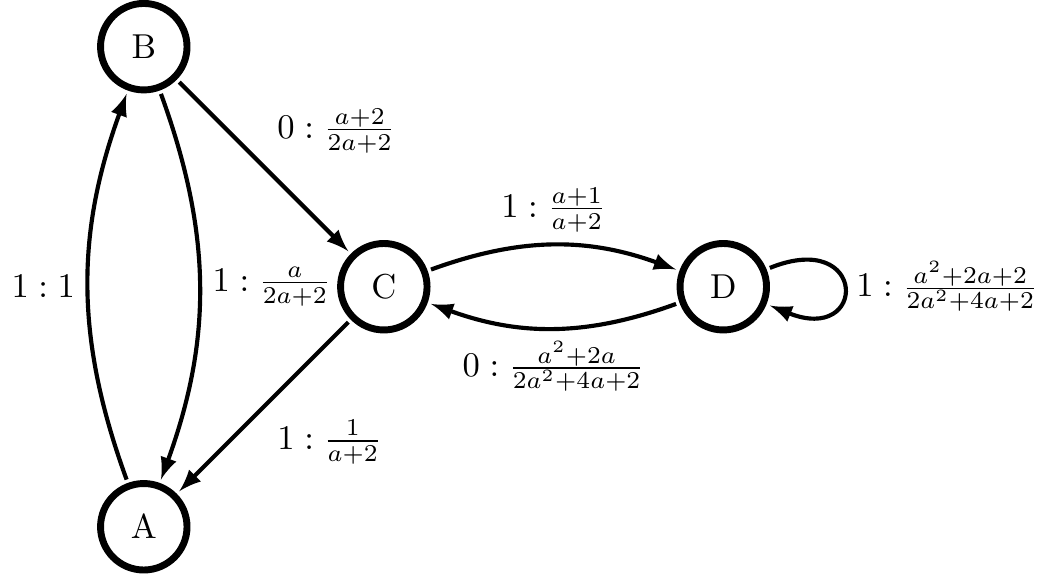}
\caption{Symbolic dynamics of the Tent Map at the Misiurewicz parameter $a$.
  (top) The map iterates points $x_n$ in the unit interval $[0,1]$ according to
  $x_{n+1} = \frac{a}{2}(1-2|x_n-\frac{1}{2}|)$, with $x_0 \in [0,1]$. The
  symbolic dynamics translates the sequence $x_0, x_1, x_2, \ldots$ of real
  values to a $0$ when $x_n\in [0,\frac{1}{2})$ and to a $1$ when $x_n\in
  [\frac{1}{2},1]$.
  (bottom) Calculations described elsewhere \cite{Jame13a} yield the \eM\ shown.
  (Reproduced from Ref. \protect{\cite{Jame13a}} with permission.)
  }
\label{fig:TentMap_setup}
\end{figure}

Up to this point, the emphasis was analyzing selected prototype infinite
Markov-order processes to illustrate the differences between CIB and OCF. In
the following, instead we apply CIB and OCF to gain insight into a nominally
more complicated process---a one-dimensional chaotic map of the unit
interval---in which we emphasize the predictive features detected. We consider
the symbolic dynamics of the Tent Map at the Misiurewicz parameter $a=\big(
\sqrt[3]{9+\sqrt{57}}+\sqrt[3]{9-\sqrt{57}} \big) 3^{-\frac{2}{3}}$, studied in
Ref. \cite{Jame13a}. Figure \ref{fig:TentMap_setup} gives both the Tent Map and
the analytically derived \eM\ for its symbolic dynamics, from there. The latter
reveals that the symbolic dynamic process is infinite-order Markov. The
bidirectional \eM\ at this parameter setting is also known. Hence, one can
directly calculate information functions as described in Sec.
\ref{sec:Examples}.

\begin{figure}[htp]
\centering
\includegraphics[width=\columnwidth]{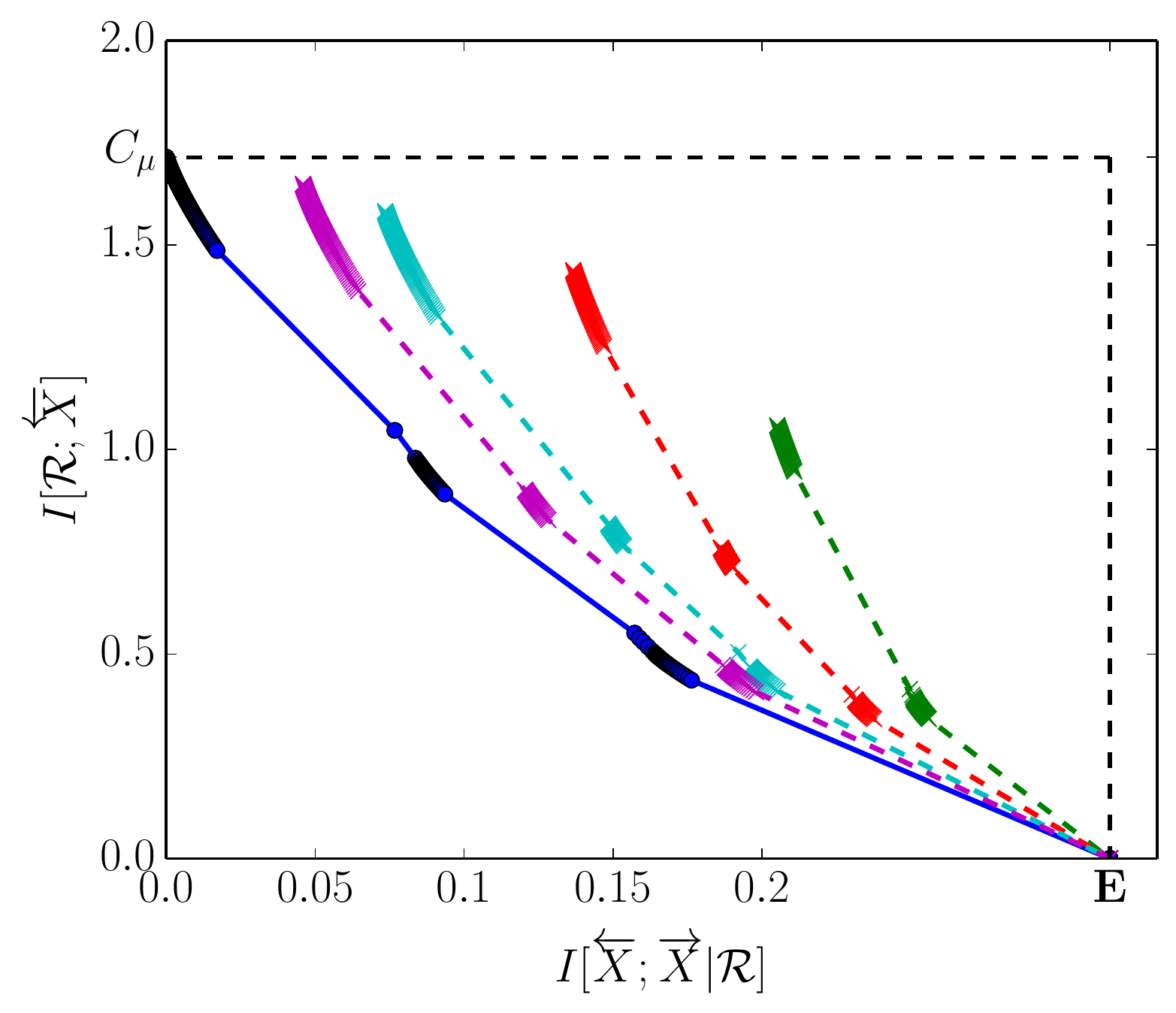}
\includegraphics[width=\columnwidth]{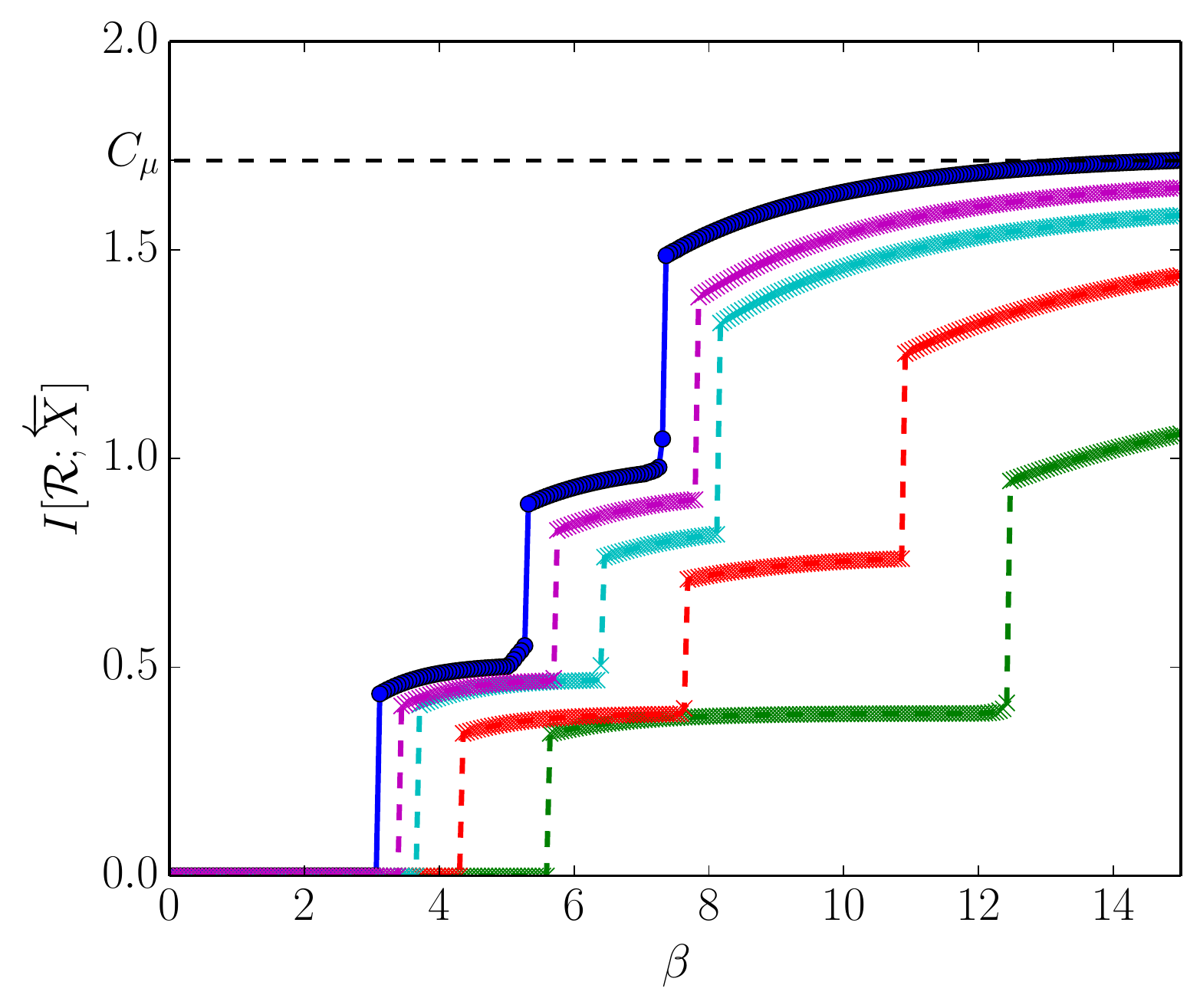}
\caption{Rate distortion analysis for symbolic dynamics of the Tent Map at the
  Misiurewicz parameter $a$ given in the text. (top) Information functions.
  (bottom) Feature curves. Comparing CIB (solid line, blue circles) and OCF
  (dashed lines, colored crosses) at several values of $L$. Legend same as
  previous.
  }
\label{fig:TentMap_PIC}
\end{figure}

From Fig.~\ref{fig:TentMap_PIC}'s information functions, one easily gleans
natural coarse-grainings, scales at which there is new structure, from the
functions' steep regions. As is typically true, the steepest part of the
predictive information function is found at very low rates and high
distortions. Though the information function of Fig. \ref{fig:TentMap_PIC}(top)
is fairly smooth, the feature curve (Fig. \ref{fig:TentMap_PIC}(bottom))
reveals phase transitions where the feature space expands a lossier
causal state into two distinct representations.

\begin{figure}[htb]
\raggedright
(a) \includegraphics[width=.22\textwidth]{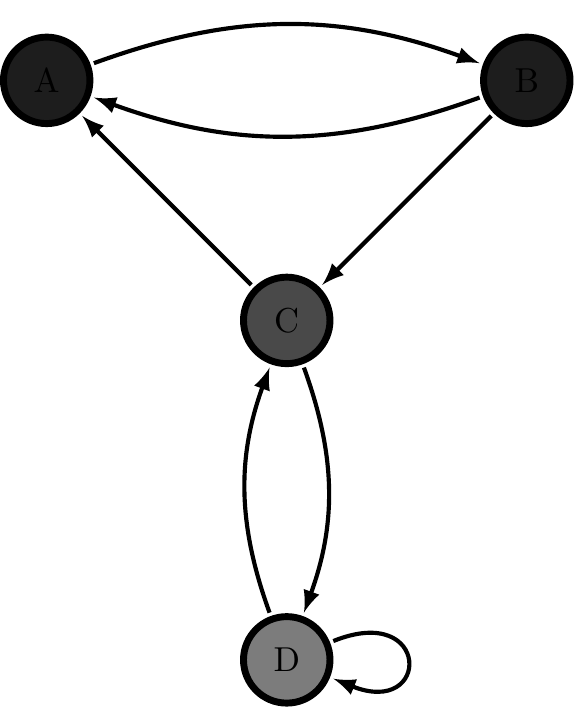} \\
(b) \includegraphics[width=.22\textwidth]{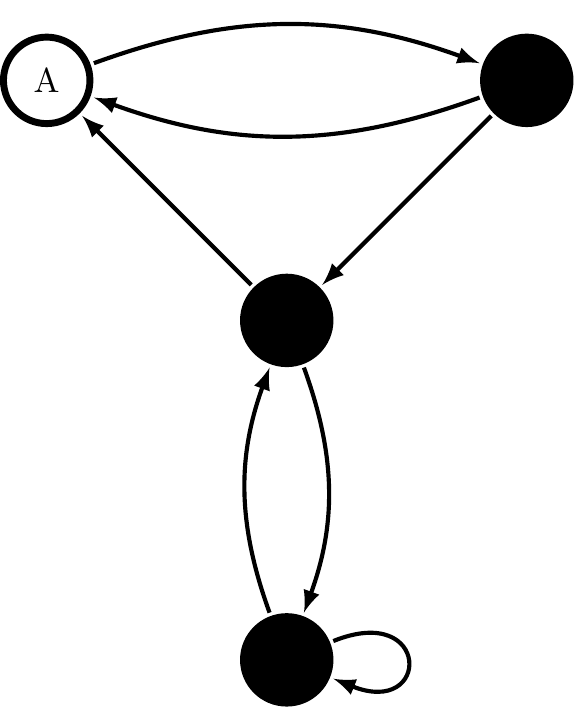}
\includegraphics[width=.22\textwidth]{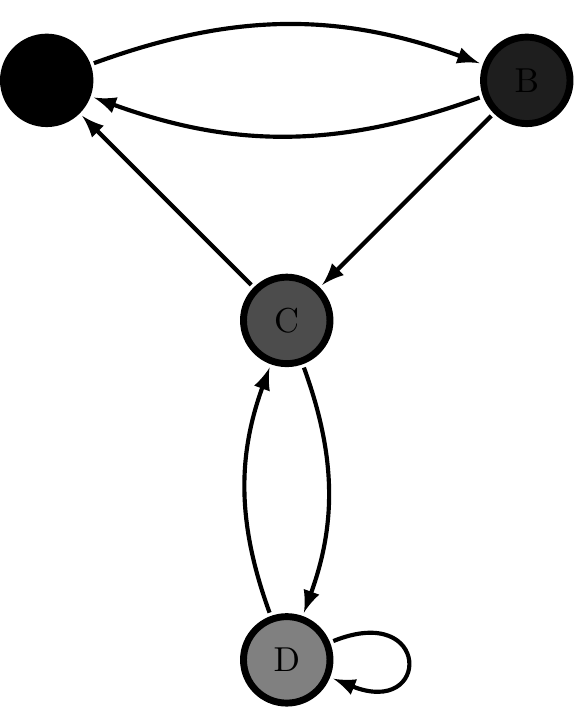} \\
(c) \includegraphics[width=.22\textwidth]{TentMap_b=2_r1}
\includegraphics[width=.22\textwidth]{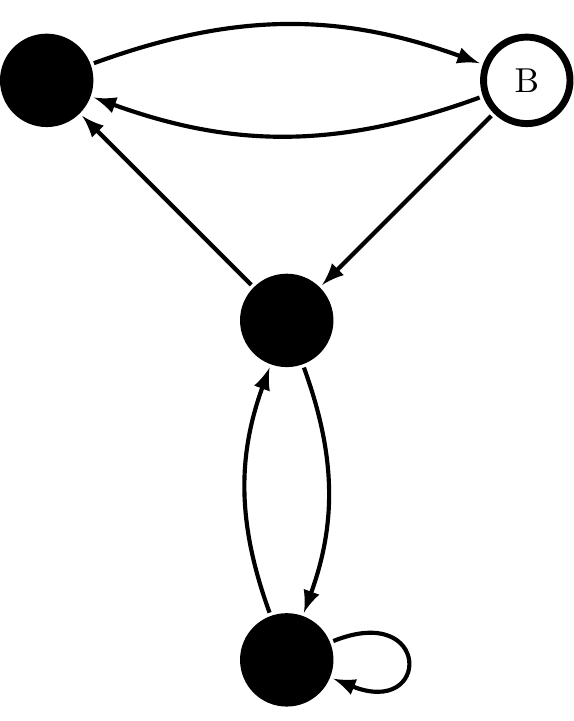}
\includegraphics[width=.22\textwidth]{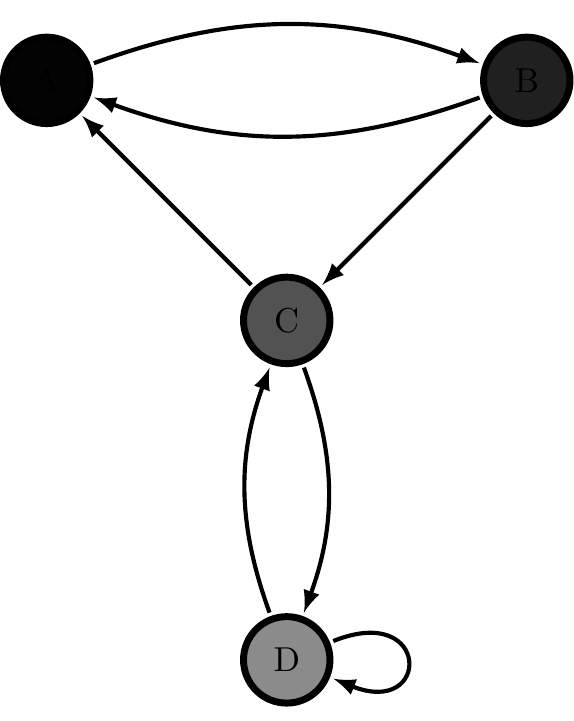} \\
(d) \includegraphics[width=.22\textwidth]{TentMap_b=2_r1}
\includegraphics[width=.22\textwidth]{TentMap_b=3_r2}
\includegraphics[width=.22\textwidth]{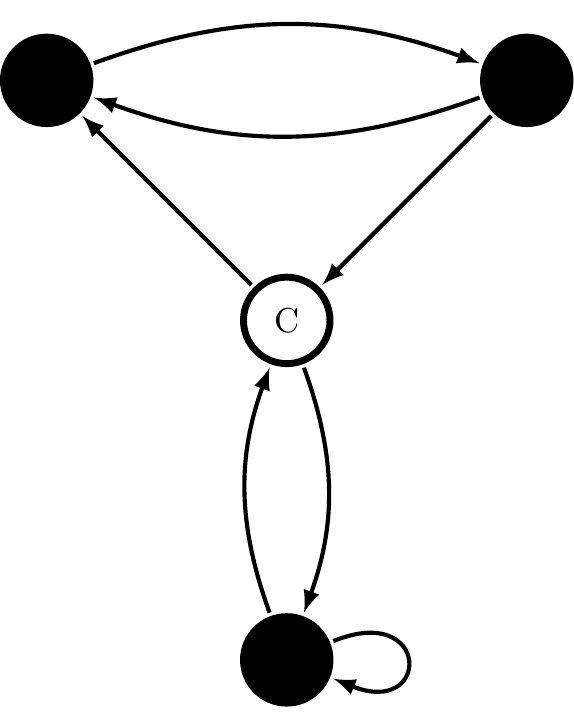}
\includegraphics[width=.22\textwidth]{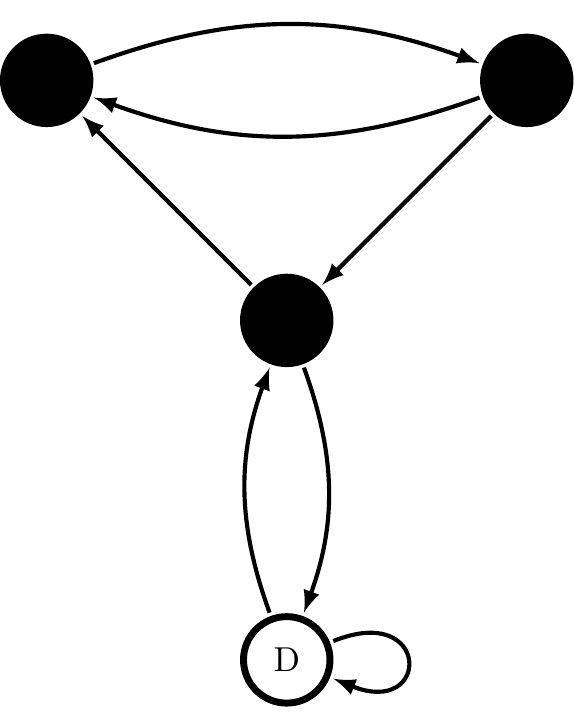} \\
\caption{Tent Map predictive features as a function of inverse temperature $\beta$:
  Each state-transition diagram shows the \eM\ in
  Fig.~\ref{fig:TentMap_setup}(bottom) with nodes gray-scaled by
  $\Prob(\St^+|\Rep=\rep)$ for each $\rep\in\Rep$. White denotes high
  probability and black low. Transitions are shown only to guide the eye.
  (a) $\beta=0.01$: one state that puts unequal weights on states $C$ and $D$.
  (b) $\beta=1.9$: two states identified, $A$ and a mixture of $C$ and $D$.
  (c) $\beta=3.1$: three states are identified, $A$, $B$, and the mixture of
  $C$ and $D$.
  (d) $\beta \to \infty$: original four states identified, $A$, $B$, $C$, and
  $D$.
  }
\label{fig:TentMapFeatureDiscover}
\end{figure}

To appreciate the changes in underlying predictive features as a function of
inverse temperature, Fig. \ref{fig:TentMapFeatureDiscover} shows the
probability distribution $\Prob(\mathcal{S}^+|\mathcal{R})$ over causal states
given each compressed variable---the features. What we learn from such phase
transitions is that some causal states are more important than others and that
the most important ones are not necessarily intuitive. As we move from lossy to
lossless ($\beta \to \infty$) predictive features, we add forward-time causal
states to the representation in the order $A$, $B$, $C$, and finally $D$. The
implication is that $A$ is more predictive than $B$, which is more predictive
than $C$, which is more predictive than $D$.  Note that this predictive
hierarchy is not the same as a ``stochastic hierarchy'' in which one prefers
causal states with smaller $\H[\MS_0|\St^+=\st^+]$. The latter is equivalent to
an ordering based on correctly predicting only one time step into the future.
Such a hierarchy privileges causal state $C$ over $B$ based on the transition
probabilities shown in Fig.~\ref{fig:TentMap_setup}(bottom), in contrast to how
CIB orders them.

\section{Computational Challenges}
\label{sec:Comp}

Proposition \ref{prop:CRD} and Thm. \ref{the:CRD} says that one can calculate PRD functions (including PIB functions) in three steps:
\begin{enumerate}
\item Theoretically derive a system's \eM\ \cite{Crut92c,Crut97a}
   using Eq. (\ref{eq:CausalEquiv}) or empirically estimate an
   \eM\ \cite{Stre13a,Stre13b,Shal02a,Pfau10a,Stil07a,CSSA,Stil13a};
\item Calculate the joint probability distribution over forward- and
	reverse-time causal states \cite{Crut92c,Crut08a,Crut08b}; and
\item Apply a rate-distortion algorithm to compress $\St^+$ to minimize
	expected distortion about $\St^-$ for a desired distortion.
\end{enumerate}
From the computational complexity viewpoint \cite{Papa94a} each step is hard.

Recall that NP refers to the class of decision problems for which a
deterministic Turing machine can verify a ``yes'' answer in polynomial-time; an
NP-hard problem is one that is as hard as the hardest problems in NP. Such
problems are quite familiar. Many computations related to spin glass Ising
models are NP-hard \cite{barahona1982computational}. Inferring an \eM\ from
samples is very likely NP-hard, given the computational complexity hardness
results on passively inferring \emph{nonprobabilistic} finite automata from only
positive examples \cite[and citations therein]{pitt1989inductive}. Calculating
$\Prob(\FSt,\PSt)$ from an \eM\ typically has run time and storage requirements
exponential in $|\FSt|$.  Solving the associated PRD optimization is known to
be NP-hard \cite{mumey2003optimal}.

It may seem as though the three-step approach is overly complicated, especially
when compared to the more common approach in which one directly clusters
finite-length pasts to retain information about finite-length futures. In point
of fact, algorithms based on clustering pasts and futures of length $L$
themselves cannot avoid the basic complexities, either. Rather, they implicitly
assume that an order-$L$ Markov model of the underlying process is sufficiently
predictive for calculating information functions. When this assumption
fails---which happens often, as Sec. \ref{sec:CurseofDimensionality}
explained---then algorithms that explicitly cluster sequences produce
suboptimal results. Recall the examples in Sec. \ref{sec:Examples}. In short,
clustering in sequence distribution space without first inferring a model
leaves one unwittingly prone to the detrimental effects model mismatch---using
sequence histograms rather than \eMs. Building predictive models first was also
found to be particularly effective when estimating a process's large deviations
\cite{Youn93a}.

\section{Conclusion}
\label{sec:Conclusion}

We introduced a new relationship between predictive rate-distortion and causal
states. Theorem $1$ of Refs. \cite{Stil07a, Stil07b} say that the predictive
information bottleneck can identify forward-time causal states, in theory. The
analyses and results in Secs.
\ref{sec:CurseofDimensionality}-\ref{sec:Examples} suggest that in practice,
when studying time series with longer-range temporal correlations, we calculate
substantially more accurate predictive information functions by deriving or
inferring an approximate \eM\ first.

The culprit is the curse of dimensionality for prediction: the number of possible sequences increases exponentially with their length. The longer-ranged the temporal correlations, the longer sequences need to be. And, as Sec. \ref{sec:CurseofDimensionality} demonstrated, a process need not have very long-ranged temporal correlations for the curse of dimensionality to rear its head. 

Section \ref{sec:elusive_info} showed that building a model may be unnecessary
if the underlying process effectively has small, finite Markov order, since
sequences are adequate proxies for a process's effective states. Sections
\ref{sec:retropredictive}-\ref{sec:tent_map}, however, then showed that when
the underlying process has long-range temporal correlations---either larger
Markov order than sequence lengths used or infinite Markov order---then
computing PRD functions directly from sequence distributions produces
quantitatively inaccurate and structurally misleading results. Since an
exhaustive survey \cite{Jame10a} shows that infinite Markov order dominates the
space of processes generated by even finite HMMs, these problems are likely
generic and could very well have affected a number of existing calculations of
rate-distortion functions, calling into question derived interpretations.

\begin{figure}[htp]
\includegraphics[width=\columnwidth]{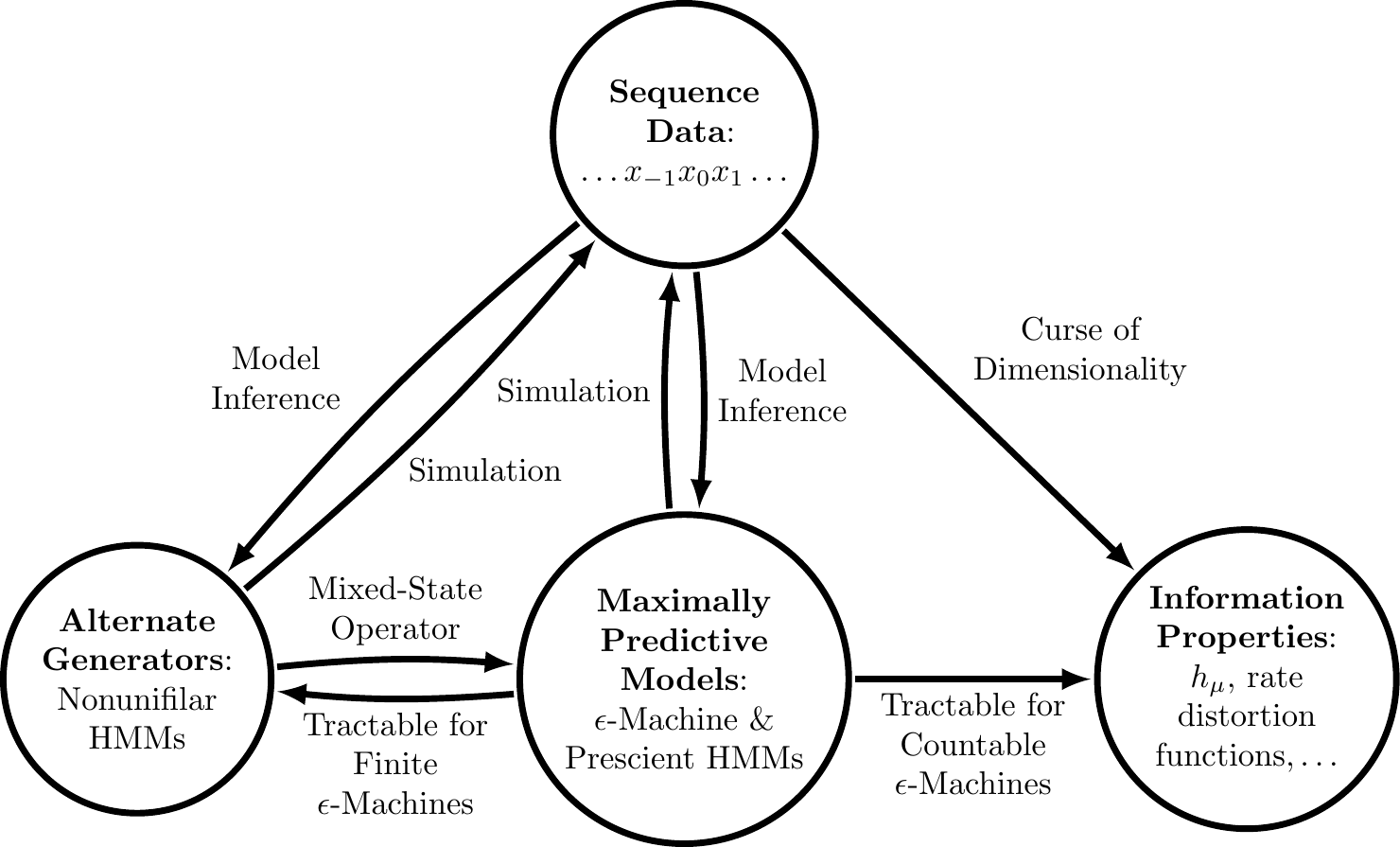}
\caption{\textbf{Prescient models and inferring information properties:}
  Estimating information measures directly from sequence data encounters a
  curse of dimensionality or, in other words, severe undersampling. Instead,
  one can calculate information measures in closed-form from (derived or
  inferred) maximally predictive (prescient) models \cite{Crut13a}.
  Rate-distortion functions are now on the list of information
  properties that can be accurately calculated.
  Alternate generative models that are \textit{not} prescient cannot be used
  directly, as Blackwell showed in the 1950s \cite{Blac57b}.
  }
\label{fig:setup}
\end{figure}

On the one hand, with these thoughts in mind, our results can be interpreted as
a cautionary tale about the curse of dimensionality inherent in predictive
rate-distortion. On the other hand, quantifying that curse of dimensionality
fully in Sec. \ref{sec:Examples} suggested new algorithms for accurately and
efficiently calculating information functions and feature curves.
Figure~\ref{fig:setup} highlights the two sides of this coin and the
alternative strategies that were compared.

These lessons echo that found when analyzing a process's large deviations
\cite{Youn93a}: Estimate a predictive model first and use it to estimate the
probability of extreme events, events that almost by definition are not in the
original data used for model inference. There may be applications that only
need temporally local predictive feature extraction---e.g., that in Ref.
\cite{Creutzig08}---and, then, by assumption there is no such curse of
dimensionality. Even then, one often assumes that temporally local learning
rules will yield temporally long-ranged predictive features. In this situation,
causal rate distortion could provide benchmarks for testing the performance of
temporally local predictive feature extractors on infinite-order Markov
processes.

The relationship between predictive rate-distortion and causal states runs much
deeper than we probed here. At a minimum, CRD is a useful theoretical tool for
analyzing coarse-grained features. One technical aspect appeared in our
discussion of how the bidirectional machine's switching maps determine much of
an information function's shape. Perhaps more importantly, though, the
predictive features identified using these methods are statistics and not full
machines with both a state space and dynamic \cite{Shal99a}. (Although one can
certainly build lossy machines from these predictive features they are likely
to be nonunifilar and not even generate the given process.) Inferring lossy
predictive machines requires a different approach, such as the recursive
information bottleneck \cite{Stil13a}. Reformulating that objective function in
terms of the bidirectional machine is an important avenue of future research.
Success in developing this will prove useful in sensorimotor loop applications,
where one learns predictive models actively rather than passively
\cite{Stil07c, singh2004predictive, dutech2013partially, singh2003learning}.

Section~\ref{sec:CRD}'s methods can be directly extended to completely
different rate-distortion settings, such as when the underlying minimal
directed acyclic graphical model between compressed and relevant random
variables is arbitrarily large and highly redundant. Also, despite the
restrictions that Prop. \ref{prop:CRD} places on the distortion measures for
which CRD and PRD are equivalent, the discussion in App. \ref{sec:App1}
suggests that CRD can be used instead of PRD if one correctly modifies the
distortion measure. This opens up a wide range of applications; for example,
those in which other properties, besides structure or prediction, are desired,
such as optimizing utility functions.

Finally, one immediate application---the construction of a predictive
hierarchy---was suggested in Sec. \ref{sec:tent_map}, using the stochastic
process defined by symbolic dynamics of a chaotic dynamical system. We saw that
rate-distortion analysis provided a principled way to determine which temporal
or spatial structures are emergent. In other words, CIB is a new tool for
accurately identifying emergent macrostates of a stochastic process
\cite{Crut92c}. Used in this way, CIB becomes a tool relevant to biological,
neurobiological, and social science phenomena in which the key emergent
features are not known a priori or from first-principles calculation. In the
context of neurobiological data, for example, such macrostates can provide
approximately predictive models of neural spike trains; e.g., see Ref.
\cite{Shaliz10a}. In the context of social science data, they might be related
to new kinds of community organization. While it is encouraging to look
forward, we appreciate that natural processes are quite complicated and that
there is quite a way to go before we have fully automated detection of emergent
macrostates.

\section*{Acknowledgments}

The authors thank C. Ellison, C. Hillar, I. Nemenman, and S. Still for helpful
discussions and the Santa Fe Institute for its hospitality during visits. JPC
is an SFI External Faculty member. This material is based upon work supported
by, or in part by, the U. S. Army Research Laboratory and the U. S. Army
Research Office under contract number W911NF-13-1-0390. SM was funded by a
National Science Foundation Graduate Student Research Fellowship and the U.C.
Berkeley Chancellor's Fellowship.

\appendix 

\section{Causal Rate-Distortion Proofs}
\label{sec:App1}

\newcommand{\CSSet} { \CausalStateSet } 
\newcommand{\CSCBook} { \Rep_{\epsilon^+} } 
\newcommand{\CSCWord} { \rep_{\epsilon^+} } 
\newcommand{\HistCBook} { \Repa_{\past} } 
\newcommand{\HistWord} { \repa_{\past} } 
\newcommand{\AltHistCBook} { \Repb_{\past} } 

We assume familiarity with rate-distortion theory on the level of Refs.
\cite[Ch. 8]{Yeun08a} and \cite{Gray90b}.

We designed the proofs to be as elementary as possible; they are basically
repeated applications of the Markov loop in Fig.~\ref{fig:Intuition1} and the
pullback method from probability theory. They are also in effect sketches more
than they are proofs. This allows us to highlight their construction nature.
All statements can \emph{almost} be proven through repeated applications of the
Markov loop in Fig.~\ref{fig:Intuition1} to the formal solution for optimal
stochastic codebooks given by Theorem $4$ of Ref. \cite{Tish00a}. The
``almost'' comes from the fact that not all solutions to the original
rate-distortion objective function maximize the annealing objective function
\cite{parker2010symmetry}.

Theorem \ref{the:CRD} does not necessarily mean that the solutions to the
annealing objectives of PIB and CIB are the same. Rather, it only finds
equivalence between solutions to the original rate distortion optimizations.
However, the same arguments straightforwardly imply that the annealing objective
functions associated with PIB and CIB also have equivalent solutions.

We consider two types of information sources and corresponding encoders. For
one, the input information source is $\Past$---the process in question---with
sample space $\SampleSpace$. Its associated codebook is denoted by $\Rep$ with
sample space $\SampleSpace$. (The sample space is specified for concreteness,
though unnecessary in general.) And, the distortion measure is $d: \SampleSpace
\times \SampleSpace \rightarrow \Re^+$.

For the other, the input information source is $\St^+$---the
internal causal-state process---with sample space $\epsilon^+(\SampleSpace)$
determined by the causal-state map;
see Ref. \cite{Shal98a}. Its codebook is denoted $\Repa$ with sample space
$\epsilon^+ \big(\SampleSpace \big)$. (Again, this is for concreteness, but
unnecessary in general.) And, its distortion measure is $\widehat{d}:\epsilon^+
\big(\SampleSpace \big) \times \epsilon^+ \big(\SampleSpace \big) \rightarrow
\Re^+$.

Implicitly, by specifying that our codebooks produce codewords in the same
sample space of the inputs themselves, we combine the encoder and decoder of
Fig. \ref{fig:Intuition2}(top) into a single information processing unit.

Using the fact that the two source sample spaces and codebooks are intimately
related via the causal-state map $\st^+ = \epsilon^+(\past)$, we construct codebooks in one sample space
out of codebooks from the other. For instance, a given process codebook $\Rep$
implies a causal-state codebook $\CSCBook$ with realizations $\CSCWord \in
\epsilon^+(\SampleSpace)$:
\begin{align*}
& \Prob (\CSCBook |\St^+=\st^+) \\
  & = \sum_{\past\in\SampleSpace}
  \Prob \big(\CSCBook | \Past=\past,\St^+=\st^+ \big)
  \Prob \big(\Past=\past|\St^+=\st^+ \big) \\
  & = \sum_{\past:\epsilon^+(\past)=\st^+} \Prob \big(\Rep|\Past=\past \big)
  \Prob \big(\Past=\past \big)
  ~.
\end{align*}
This is essentially the pullback method of probability theory.
Similarly, a given causal-state codebook $\Repa$ can be naturally extended to a
process codebook $\HistCBook$ with realizations $\HistWord \in \AllPasts$:
\begin{align*}
\Prob \big(\HistCBook|\Past=\past \big)
  & = \Prob \big(\Repa|\St^+=\epsilon^+(\past) \big)
  ~.
\end{align*}
In this way, we compare $\Repa$ to $\Rep$, using $\HistCBook$. To
simplify these comparisons, if $\mathcal{Q}$ is some codebook, in the following
$\mathcal{Q}_{\epsilon^+}$ and $(\mathcal{Q})_{\epsilon^+}$ denote a codebook
over causal states and $\mathcal{Q}_{\past}$ and $(\mathcal{Q})_{\past}$ a
codebook over process pasts.

Here, we only consider distortion measures of the form:
\begin{align*}
d(\past,\rep) & = f \big(\Prob(\Future|\Past=\past),\Prob(\Future|\Rep=\rep)
\big)
\end{align*}
and
\begin{align*}
\widehat{d}(\st^+,\repa) & = f
\big(\Prob(\Future|\St^+=\st^+),\Prob(\Future|\Repa=\repa) \big)
  ~,
\end{align*}
for some unspecified $f(\cdot,\cdot)$ that quantifies differences between probability
distributions; e.g., Kullback-Liebler or Shannon-Jensen divergence. Causal
shielding implies that:
\begin{align*}
\Prob(\Future|\Past=\past) = \Prob(\Future|\St^+=\epsilon^+(\past))
\end{align*}
and:
\begin{align*}
\Prob(\Future|\Rep)
  & = \sum_{\st^+ \in \CSSet^+} \Prob(\Future|\St^+=\st^+)\Prob(\St^+=\st^+|\Rep) \\
  & = \sum_{\st^+ \in \CSSet^+}
  \Prob(\Future|\St^+=\st^+)\Prob(\St^+=\st^+|\CSCBook)
  ~.
\end{align*}
By construction, then:
\begin{align*}
d(\past,\rep) = \widehat{d}(\epsilon^+(\past),\CSCWord)
\end{align*}
and
\begin{align*}
\widehat{d}(\st^+,\repa) = d(\past,\HistWord)
  ~,
\end{align*}
for all $\past$ such that $\epsilon^+(\past)=\st^+$.

On the one hand, when the information source is $\Past$, we consider the
following rate-distortion function:
\begin{align*}
R(D) & = \min_{\langle d(\past,\rep)\rangle_{\Rep,\Past}\leq D} \I[\Rep;\Past]
\end{align*}
and its associated optimal codebook:
\begin{align*}
\Rep^*_D & = \argmin_{\langle d(\past,\rep)\rangle_{\Rep,\Past}\leq D}
  \I[\Rep;\Past]
  ~.
\end{align*}
On the other, when the information source is $\St^+$, we consider the following
rate-distortion function:
\begin{align*}
\widehat{R}(D) & = \min_{\langle \widehat{d}(\st^+,\repa)\rangle_{\Repa,\St^+}\leq D}
\I[\Repa;\St^+]
\end{align*}
and its associated optimal codebook:
\begin{align}
\Repa^*_D & = \argmin_{\langle \widehat{d}(\st^+,\repa)\rangle_{\Repa,\St^+}\leq D}
\I[\Repa;\St^+]
  ~.
\label{eq:App1}
\end{align}
The lemma below, a more precise version of the main text's Lemma
\ref{lem:1}, states their relationship.

\vspace{0.1in}
{\noindent \textbf{Lemma \ref{lem:1}.}
$R(D) = \widehat{R}(D)$ and $\Rep^*_D \simeq \Repa^*_D$ for all achievable $D$.
}


{\ProLem First, to establish the isomorphism we construct a map between the
process and causal-state codebooks, showing that $\Rep^*_D = ((\CSCBook)^*_D)_{\past}$ via a proof by contradiction. Suppose they are not equal.
As described earlier:
\begin{align*}
D & = \big\langle d(\past,\rep) \big\rangle_{\Rep^*_D,\Past} \\
  & = \big\langle \widehat{d} (\st^+,\CSCWord)
  \big\rangle_{(\CSCBook)^*_D,\St^+} \\
  & = \big\langle d(\past,(\CSCWord)_{\past})
   \big\rangle_{ ((\CSCBook)^*_D)_{\past} , \Past }
  ~.
\end{align*}
So, the two codebook random variables have the same expected distortion. Since $\Rep^*_D \neq ((\CSCBook)^*_D)_{\past}$,
$\Prob(\Rep^*_D,\Past|\St^+) \neq \Prob(\Rep^*_D|\St^+)\Prob(\Past|\St^+)$ and
thus $\I[\Rep^*_D;\Past|\St^+] > 0$. This implies that
$((\CSCBook)^*)_\past$ has a lower coding cost: 
\begin{align*}
\I[\Rep^*_D;\Past] & = \I[\Rep^*_D;\St^+] + \I[\Rep^*_D;\Past|\St^+] \\
  & > \I[\Rep^*_D;\St^+]\\
  & = \I[(\CSCBook)^*_D;\St^+] \\
  & = \I[((\CSCBook)^*_D)_\past;\St^+]
  ~.
\end{align*}
Since $((\CSCBook)^*_D)_\past$ is a codebook
with the same expected distortion as $\Rep^*_D$ and lower coding cost, our
assumption that they are not equal is incorrect and so
$\Rep^*_D = ((\CSCBook)^*_D)_\past$.

Second, to show that $(\HistCBook)^*_D = \Rep^*_D$, we only need to show that
$(\HistCBook)^*_D = (\CSCBook)^*_D$ since the extension from
codebooks $\Repa$ to codebooks $\Rep$ is injective. The reduced codebook
$(\CSCBook)^*_D$ inherits a constraint:
\begin{align*}
\langle d(\past,\rep)\rangle_{\Rep^*_D,\Past} = \langle
\widehat{d}(\st^+,\CSCWord)\rangle_{(\CSCBook)^*_D,\St^+} \leq D
\end{align*}
and must minimize coding cost $\I[\Rep^*_D;\Past] = \I[(\CSCBook)^*_D;\St^+]$,
as described above. In short, $(\CSCBook)^*_D$ satisfies:
\begin{align}
(\CSCBook)^*_D & = \argmin_{\langle \widehat{d}(\st^+,\CSCWord)
  \rangle_{(\CSCBook)^*_D,\St^+} \leq D}
  \I[(\CSCBook)^*_D;\St^+]
  ~.
\label{eq:App2}
\end{align}
Comparing Eq.~(\ref{eq:App1}) to Eq.~(\ref{eq:App2}) yields:
\begin{align*}
\Repa^*_D & = (\CSCBook)^*_D \\
  (\Repa^*_D)_\past & = ((\CSCBook)^*_D)_\past \\
  & = \Rep^*_D ~,
\end{align*}
as desired. Using earlier manipulations, we also see that:
\begin{align*}
R(D) & = \I[\Rep^*_D;\Past] \\
  & = \I[(\CSCBook)^*_D;\St^+] \\
  & = \I[\Repa^*_D;\St^+] \\
  & = \widehat{R}(D)
  ~.
\end{align*}
completing the proof sketch.
}

Next, consider an alternate class of causal-state codebooks $\Repb$ and
realizations $\repb$ for coding information sources $\St^+$. These are
distinguished by a new distortion measure:
\begin{align*}
\widetilde{d}(\st^+,\repb)
  = f \big(\Prob(\St^-|\St^+=\st^+),\Prob(\St^-|\Repb=\repb) \big)
  ~.
\end{align*}
The associated rate-distortion function is:
\begin{align*}
\widetilde{R}(D)
  = \min_{\langle \widetilde{d}(\st^+,\repb)
  \rangle_{\Repb,\St^+}\leq D} \I[\Repb;\St^+]
  ~,
\end{align*}
with optimal codebooks:
\begin{equation}
\Repb^*_D = \argmin_{\langle \widetilde{d}(\st^+,\repb)
  \rangle_{\Repb,\St^+}\leq D} \I[\Repb;\St^+]
  ~.
\end{equation}
We would like to know the relationship between $R(D)$ and $\widetilde{R}(D)$,
among other things. The proposition below, a more precise version of the main
text's Prop. \ref{prop:CRD}, states their relationship.

\vspace{0.1in}
{\noindent \textbf{Proposition \ref{prop:CRD}.}
$R(D) = \widetilde{R}(D)$ and $(\Repb^*_D)_\past = \Rep^*_D$ for all
achievable $D$ \emph{if}:
\begin{align*}
f \big(\Prob(\Future & |\St^+=\st^+),\Prob(\Future|\Repb=\repb) \big) \\
  & = f \big(\Prob(\St^-|\St^+=\st^+),\Prob(\St^-|\Repb=\repb) \big)
  ~.
\end{align*}
}

{\ProProp Given the condition,
$\widehat{d}(\st^+,\repb) = \widetilde{d}(\st^+,\repb)$
and $\widehat{d}$ and $\tilde{d}$ are equivalent distortion measures. Then
$\Repa^*_D = \Repb^*_D$ and $\widehat{R}(D) = \widetilde{R}(D)$.
The conclusion follows from Lemma \ref{lem:1} above.
}

There are many distortion measures that do not satisfy the constraints in
Prop.~\ref{prop:CRD}.  As an example, we focus on mean squared error and still can use the Markov chains $\Rep \rightarrow\St^-\rightarrow\Future$ and $\Past\rightarrow\St^-\rightarrow\Future$ to some benefit:
\begin{widetext}
\begin{eqnarray}
d(\past,\rep) &=& \sum_{\future} \left(\Prob(\Future=\future|\Past=\past) - \Prob(\Future=\future|\Rep=\rep)\right)^2 \\
&=& \sum_{\st^-} \left(\sum_{\future\in\epsilon^-(\st^-)} \Prob(\Future=\future|\St^-=\st^-)^2\right) \left(\Prob(\St^-=\st^-|\Past=\past) - \Prob(\St^-=\st^-|\Rep=\rep)\right)^2.
\end{eqnarray}
\end{widetext}
Here, each reverse-time causal state is associated with a weighting factor:
\begin{align*}
\sum_{\future\in\epsilon^-(\st^-)} \Prob(\Future=\future|\St^-=\st^-)^2
  ~.
\end{align*}
This weighting factor typically vanishes if the futures are semi-infinite,
since:
\begin{align*}
\sum_{\future\in\epsilon^-(\st^-)} \Prob(\Future=\future|\St^-=\st^-) = 1
\end{align*}
and
$0\leq \Prob(\Future=\future|\St^-=\st^-)$. This issue is easily addressed by
rescaling the weighting factor for finite-length futures by an appropriate
function of the length and then taking the limit of this rescaled weighting
factor as the length tends to infinity.

So, again, distortion measures that do not satisfy Prop. \ref{prop:CRD}'s conditions implicitly privilege one reverse-time causal state over another.

Finally, we concentrate on a particularly useful distortion measure that satisfies the above constraint. OCI compresses $\Past$ to retain
information about $\Future$, resulting in the information function:
\begin{align*}
R(I_0) & = \min_{\I[\Rep;\Future]\geq I_0} \I[\Past;\Rep]
\end{align*}
and the associated representation is:
\begin{align*}
\Rep^*_{I_0} & = \argmin_{\I[\Rep;\Future]\geq I_0} \I[\Past;\Rep]
  ~.
\end{align*}
In contrast, CIB, using causal states as proxies for the past and future,
compresses $\St^+$ to retain information about $\St^-$, yielding the information function:
\begin{align*}
\widetilde{R}(I_0) & = \min_{\I[\Repb;\St^-]\geq I_0} \I[\St^+;\Repb]
  ~,
\end{align*}
with associated representations:
\begin{align*}
\Repb^*_{I_0} & = \argmin_{\I[\Repb;\St^-]\geq I_0} \I[\St^+;\Repb]
  ~.
\end{align*}
OCI maximizes \emph{information} about the future, whereas PRD minimizes
\emph{distortion} of predictions of the future. To emphasize the difference, we
replaced $D$ with $I_0$, following Ref. \cite{parker2010symmetry}.

\vspace{0.1in}
{\noindent \textbf{Theorem \ref{the:CRD}.}
$R(I_0) = \widetilde{R}(I_0)$ and $(\Repb^*_{I_0})_\past = \Rep^*_{I_0}$
for all achievable $I_0$.
}

{\ProThe First, consider PRD with distortion measure $d(\past,\rep) =
\DKL[\Prob(\Future|\Past=\past) || \Prob(\Future|\Rep=\rep)]$. Recall that
the reverse-time causal states ``causally shield'' the past from future; just
as the forward-time causal states do via Markov chain $\Past \rightarrow
\St^- \rightarrow \Future$. Then:
\begin{widetext}
\begin{align*}
\Prob(\Future = \future|\Past=\past)
  & = \sum_{\st^-} \Prob(\Future=\future|\St^-=\st^-,\Past=\past)
  \Prob(\St^-=\st^-|\Past=\past) \\
  & = \Prob(\Future=\future|\St^-=\epsilon^-(\future))
  \Prob(\St^-=\epsilon^-(\future)|\Past=\past)
  ~.
\end{align*}
Similarly, since $\Rep \rightarrow \St^- \rightarrow \Future$:
\begin{align*}
\Prob( \Future=\future|\Rep=\rep)
  = \Prob(\Future=\future|\St^-=\epsilon^-(\future))
  \Prob(\St^-=\epsilon^-(\future)|\Rep=\rep)
  ~.
\end{align*}

The $\DKL$ distortion measure satisfies the conditions of Prop. \ref{prop:CRD}:
\begin{align*}
d(\past,\rep) & = \DKL[\Prob(\Future|\Past=\past) || \Prob(\Future|\Rep=\rep)] \\
  & = \sum_{\future} \Prob(\Future=\future|\Past=\past) \log
  \frac{\Prob(\Future=\future|\Past=\past)}{\Prob(\Future=\future|\Rep=\rep)} \\
  & = \sum_{\future} \Prob(\Future=\future|\St^-=\epsilon^-(\future))
  \Prob(\St^-=\epsilon^-(\future)|\Past=\past)
  \log \frac{\Prob(\St^-=\epsilon^-(\future)|\Past=\past)}
  {\Prob(\St^-=\epsilon^-(\future)|\Rep=\rep)} \\
  & = \sum_{\st^-} \sum_{\future:\epsilon^-(\future)=\st^-}
  \Prob(\Future=\future|\St^-=\st^-) \Prob(\St^-=\st^-|\Past=\past)
  \log \frac{\Prob(\St^-=\st^-|\Past=\past)}{\Prob(\St^-=\st^-|\Rep=\rep)} \\
  & = \sum_{\st^-} \Prob(\St^-=\st^-|\Past=\past)
  \log \frac{\Prob(\St^-=\st^-|\Past=\past)}{\Prob(\St^-=\st^-|\Rep=\rep)} \\
  & = \DKL[\Prob(\St^-|\Past=\past) || \Prob(\St^-|\Rep=\rep)]
  ~.
\end{align*}

\end{widetext}

Thus, from Prop. \ref{prop:CRD}, $(\Repb^*_D)_\past = \Rep^*_D$ for each
achievable $D$. However, and it is straightforward to show, the expected
distortion for this measure is $\I[\Past;\Future|\Rep]$. Hence,
upper bounding the expected distortion $\langle d(\past,\rep)\rangle =
\I[\Past;\Future|\Rep]\leq D$ is equivalent to lower bounding
$\I[\Rep;\Future] \geq \I[\Past;\Future]-D$; cf. Ref. \cite{Tish00a}).
Thus, $(\Repb^*_{I_0})_\past = \Rep^*_{I_0}$ for each achievable $I_0$ and
the information functions are equivalent:
\begin{align*}
R(I_0)  & = \I[\Past;\Rep^*_{I_0}] \\
        & = \I[\Past;(\Repb^*_{I_0})_\past] \\
        & = \I[\St^+;\Repb^*_{I_0}] \\
		& = \widetilde{R}(I_0)
		~.
\end{align*}
}

Note that there are other $f(\cdot,\cdot)$ satisfying Prop. \ref{prop:CRD}'s conditions, including:
\begin{itemize}
\item $f(q_1(Y),q_2(Y)) = \sum_y |q_1(y)-q_2(y)|$ and
\item $f(q_1(Y),q_2(Y)) = \DKL[\omega_1 q_1(Y)+(1-\omega_1)q_2(Y) || \omega_2 q_1(Y) + (1-\omega_2) q_2(Y)]$ for any $0\leq\omega_1,\omega_2\leq 1$.
\end{itemize}
Again, for other distortion measures, we might find that they can be expressed
as a weighted average distortion over reverse-time causal states.

Let's close with a caveat on notation. Throughout, we cavalierly manipulated
semi-infinite pasts and futures and their conditional and joint probability
distributions---e.g., $\Prob(\Future|\Past)$. This is mathematically suspect,
since then many sums should be measure-theoretic integrals, our codebooks
seemingly have an uncountable infinity of codewords, many probabilities vanish,
and our distortion measures apparently divide $0$ by $0$. So, a more formal
treatment would instead (i) consider a series of objective functions that
compress finite-length pasts to retain information about finite-length futures
for a large number of lengths, giving finite codebooks and finite sequence
probabilities at each length, (ii) trivially adapt the proofs of Lemma
\ref{lem:1}, Prop. \ref{prop:CRD} and Thm. \ref{the:CRD} for this objective
function and a version of CIB with finite-time forward and reverse-time causal
states, and (iii) take the limit as those lengths go to infinity; e.g., as in
Ref.  \cite{Crut10d}. As long as the finite-time forward- and reverse-time
causal states limit to their infinite-length counterparts, which seems to be
the case for ergodic stationary processes but not for nonergodic processes, one
recovers Lemma  \ref{lem:1}, Prop. \ref{prop:CRD} and Thm. \ref{the:CRD}. In
the service of clarity, such care was abandoned to a shorthand, leaving the
task of an expanded measure-theoretic development to elsewhere.

\section{Optimizing the CIB Objective Function}
\label{app:CRDAlgorithm}

To make CRD's development self-contained, we explain how to derive the formal
solutions and algorithms of Eqs. (\ref{eq:CRD1})-(\ref{eq:CRD3}) from the
objective function Eq. (\ref{eq:CRDObjective}). What follows is essentially a
short review of Thm. $4$ of Ref. \cite{Tish00a} when the variable to compress
is $\St^+$ and the relevant variable is $\St^-$.

Theorem \ref{the:CRD} identifies codebooks that minimize coding rate
$I[\Rep;\St^+]$ given a lower bound on predictive power. Unsurprisingly,
optimal codebooks saturate the given bounds on predictive power
\cite{Yeun08a}.
Then, the method of Lagrange multipliers implies that optimal codebooks
maximize the annealing function:
\begin{align}
\mathcal{L}_{\beta} & = \I[\Rep;\PSt] - \beta^{-1} \I[\FSt;\Rep] \nonumber \\
  & \quad\quad\quad -\sum_{\st^+} \mu(\st^+)
  \sum_{\rep}\Prob(\Rep=\rep|\St^+=\st^+)
  ~,
  \label{eq:CIBObjective2}
\end{align}
where $\beta$ and $\mu(\st^+)$ are Lagrange multipliers.  The Lagrange
multiplier $\beta$ controls the trade-off between coding cost and predictive
power, while the Lagrange multipliers $\mu(\st^+)$ enforce normalization
constraints:
\begin{align*}
\sum_{\rep} \Prob(\Rep=\rep|\St^+=\st^+) = 1
  ~.
\end{align*}
This is a slightly different objective function than in the main text. The only
difference is that in Eq.~(\ref{eq:CRDObjective}), we search only over
$\Prob(\Rep=\rep|\St^+=\st^+)$ that are normalized. The objective function in
Eq. (\ref{eq:CIBObjective2}) explicitly enforces this constraint.

Optimal codebooks satisfy $\partial
\mathcal{L}_{\beta}/\partial\Prob(\Rep|\St^+)=0$.  For notational ease, we use
$p(\rep|\st^+)$ to denote $\Prob(\Rep=\rep|\St^+=\st^+)$ and so on.  We rewrite
$\partial\mathcal{L}_{\beta}/\partial p(\rep|\st^+) = 0$ in a more useful way.
The first step is to rewrite $\I[\Rep;\PSt] = \H[\Rep] - \H[\Rep|\PSt]$ and $\I[\FSt;\Rep] = \H[\FSt] - \H[\FSt|\Rep]$. Also:
\begin{align*}
\frac{\partial}{\partial p(\rep|\st^+)}
  \sum_{\st^+} \mu(\st^+) \sum_{\rep} p(\rep|\st^+) = \mu(\st^+)
  ~.
\end{align*}
We combine these simple manipulations to obtain:
\begin{align}
0 = (1-\beta^{-1}) & \frac{\partial H[\Rep]}{\partial
p(\rep|\st^+)}-\frac{\partial H[\Rep|\PSt]}{\partial p(\rep|\st^+)} \nonumber
\\
  & + \beta^{-1} \frac{\partial H[\Rep|\FSt]}{\partial p(\rep|\st^+)}
  - \mu(\st^+)
  ~.
\label{eq:AppB1}
\end{align}
The expression $\partial \H[\Rep|\FSt] / \partial p(\rep|\st^+)$ is
straightforward to calculate, since:
\begin{align*}
\H[\Rep|\FSt] = -\sum_{\st^+} p(\st^+) \sum_{\rep} p(\rep|\st^+) \log p(\rep|\st^+)
\end{align*}
and $\partial p(\rep'|\st^+) / \partial p(\rep|\st^+) = \delta_{\rep,\rep'}$:
\begin{align}
\frac{\partial H[\Rep|\FSt]}{\partial p(\rep|\st^+)} & = - p(\st^+) (\log p(\rep|\st^+) + 1)
  ~.
\label{eq:AppB2}
\end{align}
Next, to determine $\partial \H[\Rep] / \partial p(\rep|\st^+)$, we note that $p(\rep) = \sum_{\st^+} p(\st^+) p(\rep|\st^+)$, so
$\partial p(\rep)  / \partial p(\rep|\st^+) = p(\st^+)$. Thus, we have:
\begin{align}
\frac{\partial \H[\Rep]}{\partial p(\rep|\st^+)} & = \frac{\partial \H[\Rep]}{\partial p(\rep)} \frac{\partial p(\rep)}{\partial p(\rep|\st^+)}\nonumber \\
&= - (\log p(\rep)+1) p(\st^+). \label{eq:AppB3}
\end{align}
Finally, to calculate $\partial \H[\Rep|\PSt] / \partial p(\rep|\st^+)$, we recall that $\Rep\rightarrow\St^+\rightarrow\St^-$ is a Markov chain. Hence:
\begin{align*}
p(\rep|\pst) = \sum_{\st^+} p(\rep|\st^+) p(\st^+|\st^-)
  ~,
\end{align*}
implying that:
\begin{align*}
\frac{\partial p(\rep|\pst)}{\partial p(\rep|\fst)} = p(\fst|\pst)
  ~.
\end{align*}
These give:
\begin{align}
\label{eq:AppB4}
\frac{\partial \H[\Rep|\PSt]}{\partial p(\rep|\st^+)}
  & = \sum_{\pst} \frac{\partial \H[\Rep|\PSt]}{\partial p(\rep|\pst)}
  \frac{\partial p(\rep|\pst)}{\partial p(\rep|\fst)} \\
& = - \sum_{\pst} p(\pst) (1+\log p(\rep|\pst)) p(\fst|\pst) \nonumber  \\
& = - p(\fst) - p(\fst) \sum_{\pst} p(\pst|\fst) \log p(\rep|\pst)
  ~.
  \nonumber
\end{align}
Substituting Eqs. (\ref{eq:AppB2})-(\ref{eq:AppB4}) into Eq. (\ref{eq:AppB1})
and dividing through by $p(\fst)$ yields:
\begin{align*}
\mu(\fst) &= -(1-\beta^{-1}) \log p(\rep)
  + \sum_{\pst} p(\pst|\fst) \log p(\rep|\pst) \nonumber \\
  & \quad\quad\quad - \beta^{-1} \log p(\rep|\fst)
  ~,
\end{align*}
where $\mu(\fst)$ replaced $\mu(\fst) / p(\fst)$. (Since $\mu(\fst)$ was
an unknown constant to start with, this abuse of notation is somewhat
justified.) If we recall that:
\begin{align*}
D_{KL}[p(\pst|\fst) || p(\pst|\rep)]
  & = \sum_{\pst} p(\pst|\fst) \log \frac{p(\pst|\fst)}{p(\pst|\rep)}
  ~,
\end{align*}
we can use Bayes' Rule---$p(\rep|\pst) = p(\pst|\rep) p(\rep) / p(\pst)$---and
algebra not shown here to further rewrite:
\begin{align*}
& \sum_{\pst} p(\pst|\fst) \log p(\rep|\pst)
  = -D_{KL}[p(\pst|\fst) || p(\pst|\rep)] \\
  & \quad\quad\quad + \sum_{\pst} p(\pst|\fst) \log \frac{p(\pst|\fst)}{p(\pst)}
+ \log p(\rep)
  ~.
\end{align*}
After multiplying through by $\beta$ and allowing $\mu(\fst)$ to absorb various
unimportant other constants, this implies:
\begin{align*}
\mu(\fst) & = -\log p(\rep) - \beta D_{KL}[p(\pst|\fst) || p(\pst|\rep)]
  \nonumber\\
  & \quad\quad\quad - \log p(\rep|\fst)
  ~.
\end{align*}
Finally, a slight rearrangement gives:
\begin{align}
p(\rep|\fst) & = \frac{p(\rep)}{Z(\fst)}
  e^{- \beta D_{KL}[p(\pst|\fst) || p(\pst|\rep)]}
  ~,
\label{eq:formal}
\end{align}
where $Z(\fst) = e^{-\mu(\fst)}$.  Enforcing normalization constraints,
$\sum_{\rep} p(\rep|\fst) = 1$, means that $Z(\fst)$ is a partition function.
Meanwhile, $p(\rep)$ is set implicitly via $p(\rep) = \sum_{\fst} p(\fst)
p(\rep|\fst)$.

Equation (\ref{eq:formal}) is a formal solution for the optimal codebook
$p(\rep|\fst)$. The right-hand side of this formal solution can be viewed as a
map on $p(\rep|\st^+)$, taking one codebook to a new codebook. By iterating
this map, we eventually converge to a codebook that satisfies Eq.
(\ref{eq:formal}). Equations (\ref{eq:CRD1})-(\ref{eq:CRD3}) then turn this
realization into an algorithm.  However, there are many suboptimal codebooks
for a given $\beta$ that are also extrema of $\mathcal{L}_{\beta}$. So,
practically, one is either careful about initial conditions: starting from a
variety of initial conditions and choosing the converged codebook with maximal
$\I[\Rep;\PSt]-\beta^{-1}\I[\FSt;\Rep]$, or (preferably) both.

\bibliography{chaos}

\end{document}